\documentclass[sigconf]{acmart} 
\usepackage{amsmath, amssymb, amsfonts, mathrsfs}

\usepackage{balance}
\usepackage{leading}
\usepackage{graphicx}
\usepackage{subfig}
\usepackage{textcomp}
\usepackage{multirow}
\usepackage{framed}
\usepackage{amsmath}
\usepackage{listings}
\usepackage{xspace}
\usepackage{enumitem}
\usepackage{microtype}

\RequirePackage[normalem]{ulem}
\RequirePackage{color}\definecolor{RED}{rgb}{1,0,0}\definecolor{BLUE}{rgb}{0,0,1}

\newcommand{\squishlist}{
 \begin{list}{$\bullet$}
  { \setlength{\itemsep}{0pt}
     \setlength{\parsep}{3pt}
     \setlength{\topsep}{3pt}
     \setlength{\partopsep}{0pt}
     \setlength{\leftmargin}{1.5em}
     \setlength{\labelwidth}{1em}
     \setlength{\labelsep}{0.5em} } }
	
\newcommand{\squishlisttwo}{
 \begin{list}{$\bullet$}
  { \setlength{\itemsep}{0pt}
     \setlength{\parsep}{0pt}
    \setlength{\topsep}{0pt}
    \setlength{\partopsep}{0pt}
    \setlength{\leftmargin}{2em}
    \setlength{\labelwidth}{1.5em}
    \setlength{\labelsep}{0.5em} } }

\newcommand{\squishend}{
  \end{list}  }

\newcommand{\eat}[1]{}

\newcommand{\myfigspace}[0]{-0.45cm}

\newcommand{\subsecspace}[0]{-0.20cm}

\newcommand{\eqnspace}[0]{-0.1cm}

\copyrightyear{2020}
\acmYear{2020}
\setcopyright{acmcopyright}
\acmConference[HPDC '20]{Proceedings of the 29th International Symposium on High-Performance Parallel and Distributed Computing}{June 23--26, 2020}{Stockholm, Sweden}
\acmBooktitle{Proceedings of the 29th International Symposium on High-Performance Parallel and Distributed Computing (HPDC '20), June 23--26, 2020, Stockholm, Sweden}
\acmPrice{15.00}
\acmDOI{10.1145/3369583.3392671}
\acmISBN{978-1-4503-7052-3/20/06}  

\begin{document}
\title{Modeling The Temporally Constrained Preemptions of Transient Cloud VMs}

\author{JCS Kadupitiya}
\affiliation{Indiana University}
\email{kadu@iu.edu}
\author{Vikram Jadhao}
\affiliation{Indiana University}
\email{vjadhao@iu.edu}
\author{Prateek Sharma}
\affiliation{Indiana University}
\email{prateeks@iu.edu}

\begin{abstract}
Transient cloud servers such as Amazon Spot instances, Google
Preemptible VMs, and Azure Low-priority batch VMs, can reduce cloud
computing costs by as much as $10\times$, but can be unilaterally
preempted by the cloud provider.  Understanding preemption
characteristics (such as frequency) is a key first step in minimizing
the effect of preemptions on application performance, availability,
and cost.  However, little is understood about \emph{temporally
  constrained} preemptions---wherein preemptions must occur in a given
time window.  We study temporally constrained preemptions by
conducting a large scale empirical study of Google's Preemptible VMs
(that have a \emph{maximum} lifetime of 24 hours), develop a new
preemption probability model, new model-driven resource management
policies, and implement them in a batch computing service for
scientific computing workloads.

Our statistical and experimental analysis indicates that temporally
constrained preemptions are \emph{not} uniformly distributed, but are
time-dependent and have a bathtub shape.  We find that existing
memoryless models and policies are not suitable for temporally
constrained preemptions.  We develop a new probability model for
bathtub preemptions, and analyze it through the lens of reliability
theory.  To highlight the effectiveness of our model, we develop
optimized policies for job scheduling and checkpointing.  Compared to
existing techniques, our model-based policies can reduce the
probability of job failure by more than $2\times$.  We also implement
our policies as part of a batch computing service for scientific
computing applications, which reduces cost by $5\times$ compared to
conventional cloud deployments and keeps performance overheads under
$3\%$.
\end{abstract}

\maketitle

\section{Introduction}
\label{sec:intro}

Transient cloud computing is an emerging and popular resource allocation model used by all major cloud providers, and allows unused capacity to be offered at low costs as preemptible virtual machines.
Transient VMs can be unilaterally revoked and preempted by the cloud provider, and applications running inside them face fail-stop failures.
To expand the usability and appeal of transient VMs, many systems and techniques have been proposed that  ameliorate the effects of preemptions and reduce the computing costs of applications. 
Fault-tolerance mechanisms~\cite{spotcheck, marathe2014exploiting}, resource management policies~\cite{exosphere, conductor}, and cost optimization techniques~\cite{dubois2016optispot, shastri2017hotspot} have been developed for a wide range of applications---such as interactive web services, distributed data processing, parallel computing, etc.

Transiency mitigation techniques all depend on probabilistic estimates of when and how frequently preemptions occur.
For instance, many fault-tolerance and resource optimization policies are parametrized by the mean time to failure (MTTF) of the transient VMs. 
The preemption characteristics are governed by the transient availability model chosen by the cloud provider.

\emph{Spot} markets (used by Amazon EC2's spot instances and others) are a popular model, where preemptions are governed by dynamic prices (which are in turn set using a continuous second-price auction~\cite{spot-pricing2}). 
In this paper, we focus on a different transient availability model---temporally constrained preemptions.
In this model, transient VMs have a fixed maximum lifetime, that acts as a temporal constraint on the preemption events. 
Google's Preemptible VMs are temporally constrained---they have a maximum lifetime of 24 hours, and are always preempted within the $[0,24]$ hour interval.

The temporally constrained preemption model is distinct from spot markets, and presents fundamental challenges in preemption modeling and effective use of transient VMs. 
Transiency-mitigation techniques such as VM migration~\cite{spotcheck}, checkpointing~\cite{flint, marathe2014exploiting}, diversification~\cite{exosphere}, \emph{all} use price-signals to model the availability and preemption rates of spot instances. 
With flat pricing, these approaches are not applicable. 
Furthermore, no other information about preemption characteristics is publicly available, not even coarse-grained metrics such as MTTFs. 
To address this, we develop an \emph{empirical} approach for understanding and modeling preemptions. 
We conduct a large empirical study of over 800 preemptions of Google Preemptible VMs, and develop an analytical probability model for temporally constrained preemptions. 

Due to the temporal constraint on preemptions, classical models that form the basis of preemption modeling and policies, such as memoryless exponential failure rates, are not applicable. 
We find that preemption rates are \emph{not} uniform, but bathtub shaped with multiple distinct temporal phases, and are incapable of being modeled by existing bathtub distributions such as Weibull.
We capture these characteristics by \emph{developing a new probability model}. 
Our model uses reliability theory principles to capture the 24-hour lifetime of VMs, and generalizes to VMs of different resource capacities, geographical regions, and across different temporal domains.
Using our probability model, we find that bathtub failures can reduce the recomputation overhead of preeemptions by more than  $10\times$ compared to uniform failures---which has important implications for cloud users and providers. 

We show the applicability and effectiveness of our model by developing optimized policies for job scheduling and checkpointing. 
These policies are fundamentally dependent on empirical and analytical insights from our model. 
Our job-scheduling policy uses the bathtub behavior to decide whether to run a new job on a running VM or to request a new VM, and reduces job-failure probability by $2\times$ compared to  conventional memoryless policies. 
The bathtub distribution also requires a new approach to periodic checkpointing---since existing Young-Daly~\cite{daly2006higher} checkpointing is restricted to memoryless preemptions. 
Our checkpointing policy combines our preemption model and dynamic programming to reduce the checkpointing overhead by more than $5\times$. 
These optimized policies are a building block for transient computing systems and reducing the performance degradation and costs of preemptible VMs. 
We implement and evaluate these policies as part of a batch computing service, which we also use for empirically evaluating the effectiveness of our model and policies under real-world conditions. 

Towards our goal of developing a better understanding of constrained preemptions, we make the following contributions:
\begin{enumerate} [leftmargin=12pt]

\item We conduct a large-scale, first of its kind empirical study of preemptions of Google's Preemptible VMs \footnotemark. We then show a statistical analysis of preemptions based on the VM type, temporal effects, geographical regions, etc. Our analysis 
  indicates that the 24-hour constraint is a defining characteristic, and that the preemption rates are \emph{not} uniform, but have distinct phases. 

\item We develop a probability model of constrained preemptions based on empirical and statistical insights that point to distinct failure processes underpinning the preemption rates. Our model captures the key effects resulting from the 24 hour lifetime constraint associated with these VMs, and we analyze it through the lens of reliability theory.

\item Based on our preemption model, we develop optimized policies for job scheduling and checkpointing that minimize the total time and cost of running applications. These policies reduce job running times by up to $2\times$ compared to existing preemption models used for transient VMs. 
  
\item We implement and evaluate our policies as part of a batch computing service for Google Preemptible VMs. Our service is especially suitable for scientific simulation applications, and can reduce computing costs by $5\times$ compared to conventional cloud deployments, and reduce the performance overhead of preemptible VMs to less than $3\%$. 

\end{enumerate}

\footnotetext{Preemption dataset available at https://github.com/kadupitiya/goog-preemption-data/}

\vspace*{\subsecspace}
\section{Background}

We now give an overview of transient cloud computing, and the  use of preemption models in transient computing systems. 

\vspace*{\subsecspace}
\subsection{Transient Cloud Computing}

Infrastructure as a service (IaaS) clouds such as Amazon EC2, Google Public Cloud, Microsoft Azure, etc., typically provide computational resources in the form of virtual machines (VMs), on which users can deploy their applications.
Conventionally, these VMs are leased on an ``on-demand'' basis: cloud customers can start up a VM when needed, and the cloud platform provisions and runs these VMs until they are shut-down by the customer. 
Cloud workloads, and hence the utilization of cloud platforms, shows large temporal variation. 
To satisfy user demand, cloud capacity is typically provisioned for the \emph{peak} load, and thus the average utilization tends to be low, of the order of 25\%~\cite{borg,resource-central-sosp17}. 

To increase their overall utilization, large cloud operators have begun to offer their surplus resources as low-cost servers \footnotemark with \emph{transient} availability, which can be preempted by the cloud operator at any time (after a small advance warning). 
These preemptible servers, such as Amazon Spot instances~\cite{ec2-spot}, Google Preemptible VMs~\cite{preemptible-documentation}, and Azure batch VMs~\cite{azure-batch}, have become popular in recent years due to their discounted prices, which can be 7-10$\times$ lower than conventional non-preemptible servers.
Due to their popularity among users, smaller cloud providers such as Packet~\cite{packet-spot} and Alibaba~\cite{alibaba-spot} have also started offering transient cloud servers. 

\footnotetext{We use servers and VMs interchangeably throughout the paper.}

However, effective use of transient servers is challenging for applications because of their uncertain availability~\cite{transient}. 
Preemptions are akin to fail-stop failures, and result in loss of the application memory and disk state, leading to downtimes for interactive applications such as web services, and poor throughput for batch-computing applications. 
Consequently, researchers have explored fault-tolerance techniques such as checkpointing~\cite{flint, marathe2014exploiting, spoton} and resource management techniques~\cite{exosphere} to ameliorate the effects of preemptions.
The effect of preemptions depends on the application's delay insensitivity and fault model, and mitigating preemptions for different applications remains an active research area~\cite{hourglass-eurosys19}. 

\vspace*{\subsecspace}
\subsection{Modeling Preemptions of Transient VMs}

Underlying \emph{all} techniques and systems in transient computing is the notion of using some probabilistic or even a deterministic model of  preemptions. 
Such a preemption model is then used to quantify and analyze the impact of preemptions on application performance and availability; and to design model-informed policies to minimizing the effect of preemptions. 
For example, the preemption rate or MTTF (Mean Time To Failure) of transient servers has found extensive use in selecting the appropriate type transient server for applications~\cite{exosphere, spoton}, determining the optimal checkpointing frequency~\cite{flint, marathe2014exploiting, proteus-eur17, ghit-spark-hpdc}, etc. 

Preemptions of spot market based VMs (such as EC2 spot instances) are based on their \emph{price}, which is dynamically adjusted based on the supply and demand of cloud resources. 
Spot prices are based on a continuous second-price auction, and if the spot price increases above a pre-specified maximum-price, then the server may be preempted~\cite{spot-pricing2}. 
Thus, the time-series of these spot prices can be used for understanding preemption characteristics such as the frequency of preemptions and the ``Mean Time To Failure'' (MTTF) of the spot instances. 
Publicly available~\cite{bidding4} historical spot prices have been used to characterize and model spot instance preemptions~\cite{spotcheck, bid-cloud, transient-guarantees, wolski2016providing}. 
For example, past work has analyzed spot prices and shown that the MTTFs of spot instances of different hardware configurations and geographical zones range from a few hours to a few days~\cite{wolski_probabilistic_2017, icdcs-spotlight, wolski2016providing, baughman2018predicting, wolski2017probabilistic}.
Spot instance preemptions can be modeled using \emph{memoryless}  exponential distributions~\cite{bid-cloud, hotcloud-not-bid, flint, ghit-spark-hpdc, chien-ic2e19}, which permits optimized periodic checkpointing policies such as Young-Daly~\cite{daly2006higher}. 

However, using pricing information for preemption modeling is \emph{not} a generalizable approach, and is not applicable to other models of transient availability used by other transient VMs like Google Preemptible VMs and Azure Low-priority batch VMs. 
These VMs have \emph{flat} pricing, and thus pricing cannot be used to infer preemptions. 
Moreover, these cloud providers (Google and Azure) do not expose \emph{any} public information about their preemption characteristics, even coarse grained metrics like MTTF that can be useful in mitigating preemptions~\cite{chien-ic2e19}. 
In this paper, we propose an empirical approach for modeling preemptions of temporally constrained VMs such as Google Preemptible VMs.
Our empirical data and the resulting preemption model allows the development of preemption mitigation policies. 
Google Preemptible VMs have a maximum lifetime of 24 hours, and this \emph{constrained} preemption is not memoryless, and requires new fundamental modeling approaches. 

\section{Understanding temporally constrained VM preemptions}
\label{sec:failmodel}

In our quest to understand temporally constrained preemptions, we  conduct an empirical study of preemptions of Google Preemptible VMs.
Based on our observations and insights from the study, we then develop a probability model for temporally constrained preemptions, which we later use to develop preemption-mitigating resource management application policies. 

\vspace*{\subsecspace}
\subsection{Empirical Study Of Preemptions}\label{sec:empirical}

\begin{figure}
  \includegraphics[width=0.47\textwidth]{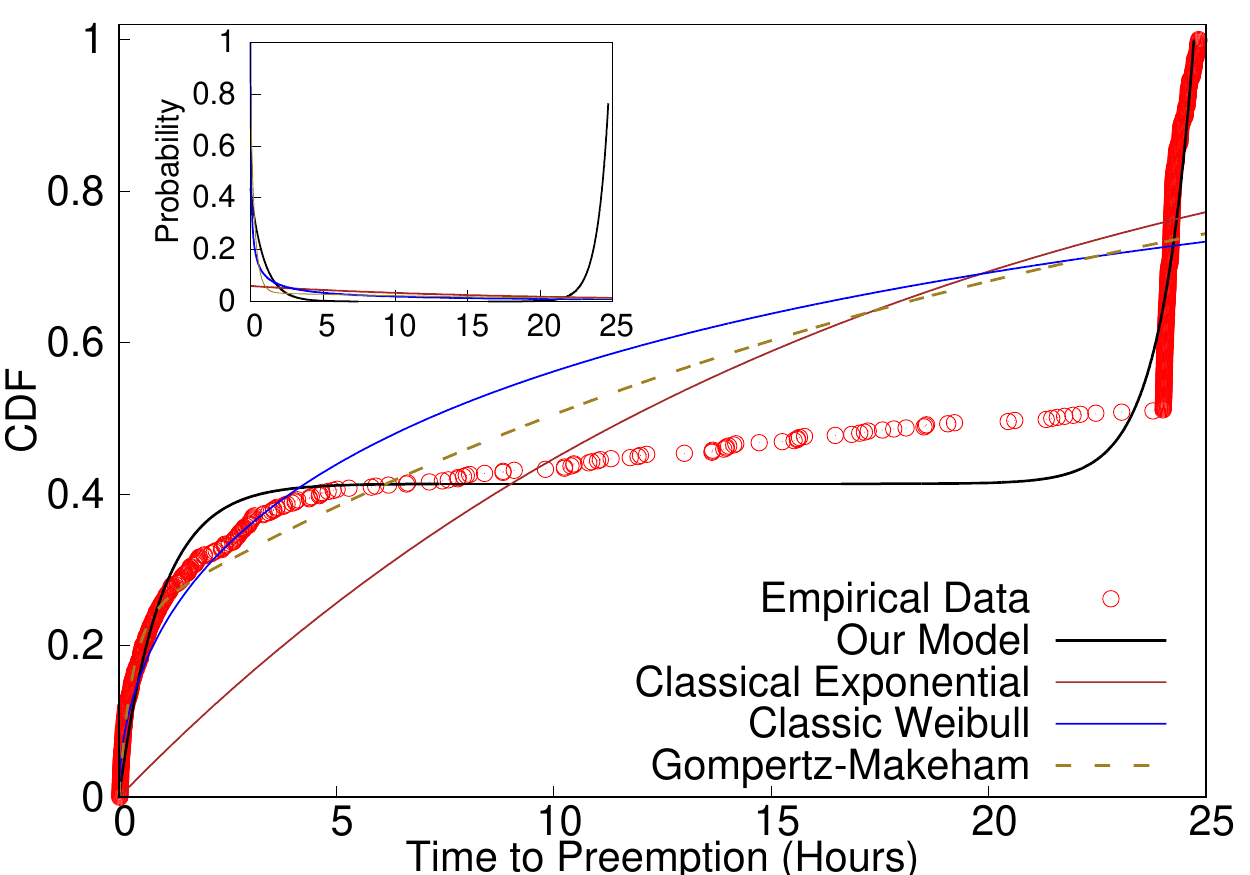} 
  \caption{CDF of lifetimes of Google Preemptible VMs. Our proposed distribution for modeling the constrained preemption dynamics provides a better fit to the empirical data compared to other failure distributions. Inset shows the probability density functions.}
  \label{fig:gcp1}
\end{figure}

\noindent \textbf{Methodology.}
We launched  870 Google Preemptible VMs of different types over a two month period (Feb--April 2019), and measured their time to preemption (i.e., their useful lifetime).
VMs of different resource capacities were launched in a four geographical regions; during days and nights and all days of the week; and running different workloads\footnotemark. 
We launched VMs in their default resource configurations (CPU and memory), and do not use custom VM sizes.
To ensure the generality of our empirical observations, VMs were not launched during well-known peak utilization days (such as Black Friday).
The preemption data collection was bootstrapped: a small amount of data points were used to estimate and model the preemption CDF, which we then used to run our batch computing service (described and evaluated in Sections~\ref{sec:impl} and \ref{sec:eval}), which generated the rest of the preemption data.  
Due to the relatively high preemption rates compared to EC2 spot instances, we were able to collect these data points for less than \$5,000. 

\footnotetext{Preemption rates can also be affected by number of VMs launched simultaneously, which we limited to between $1$ and $10$.}

A sample of over 100 such preemption events are shown in Figure~\ref{fig:gcp1}, which shows cumulative distribution function (CDF) of the lifetime of the \texttt{n1-highcpu-16} VM in the \texttt{us-east1-b} zone. 
Our empirical approach allows us to make the following  observations.

\begin{figure*}
  \subfloat[Preemption characteristics of different VM types. Larger VMs are more likely to be preempted.
  \label{fig:cdf-comparison}]
  {  \includegraphics[width=0.3\textwidth]{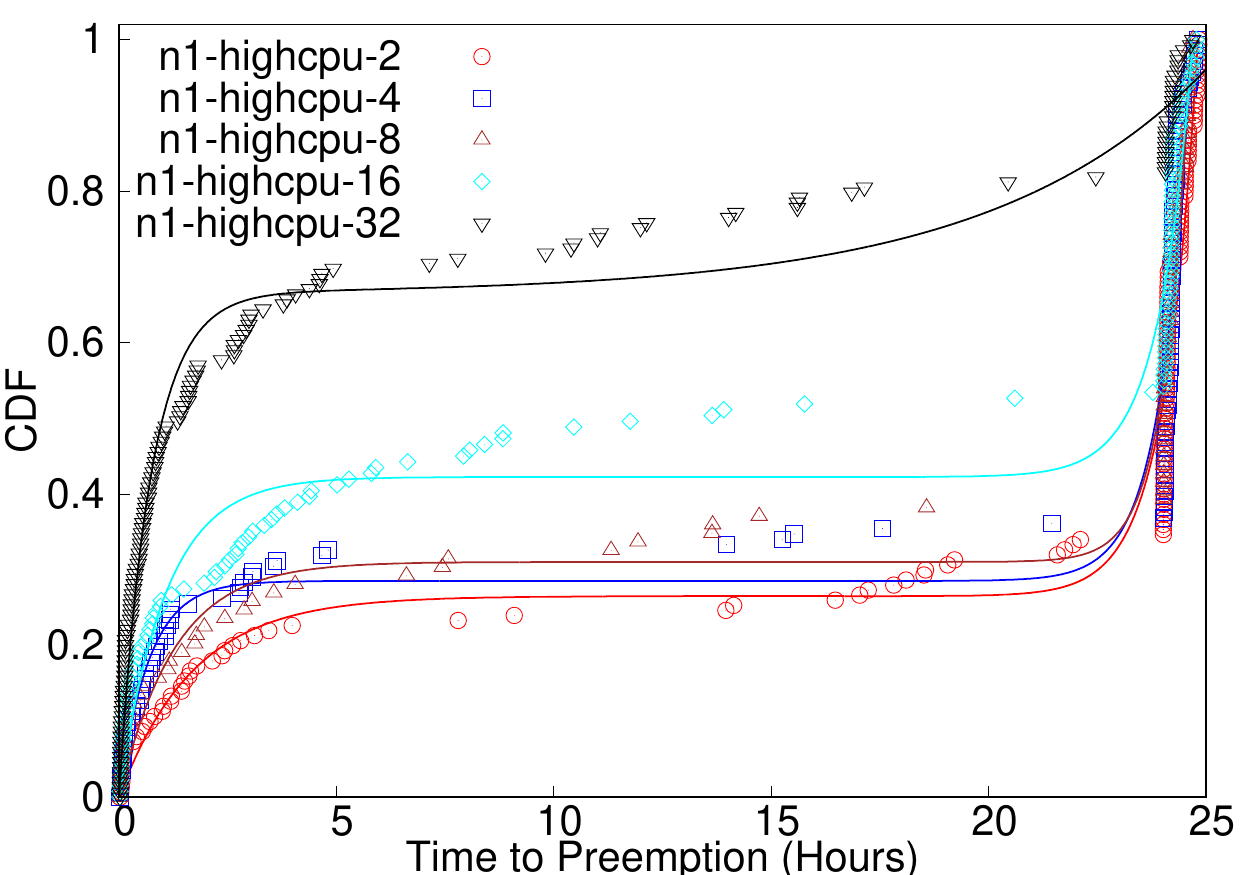} }
  \hfill
\subfloat[Variations due to time of day and workload. \label{fig:time-breakdown}]
{  \includegraphics[width=0.3\textwidth]{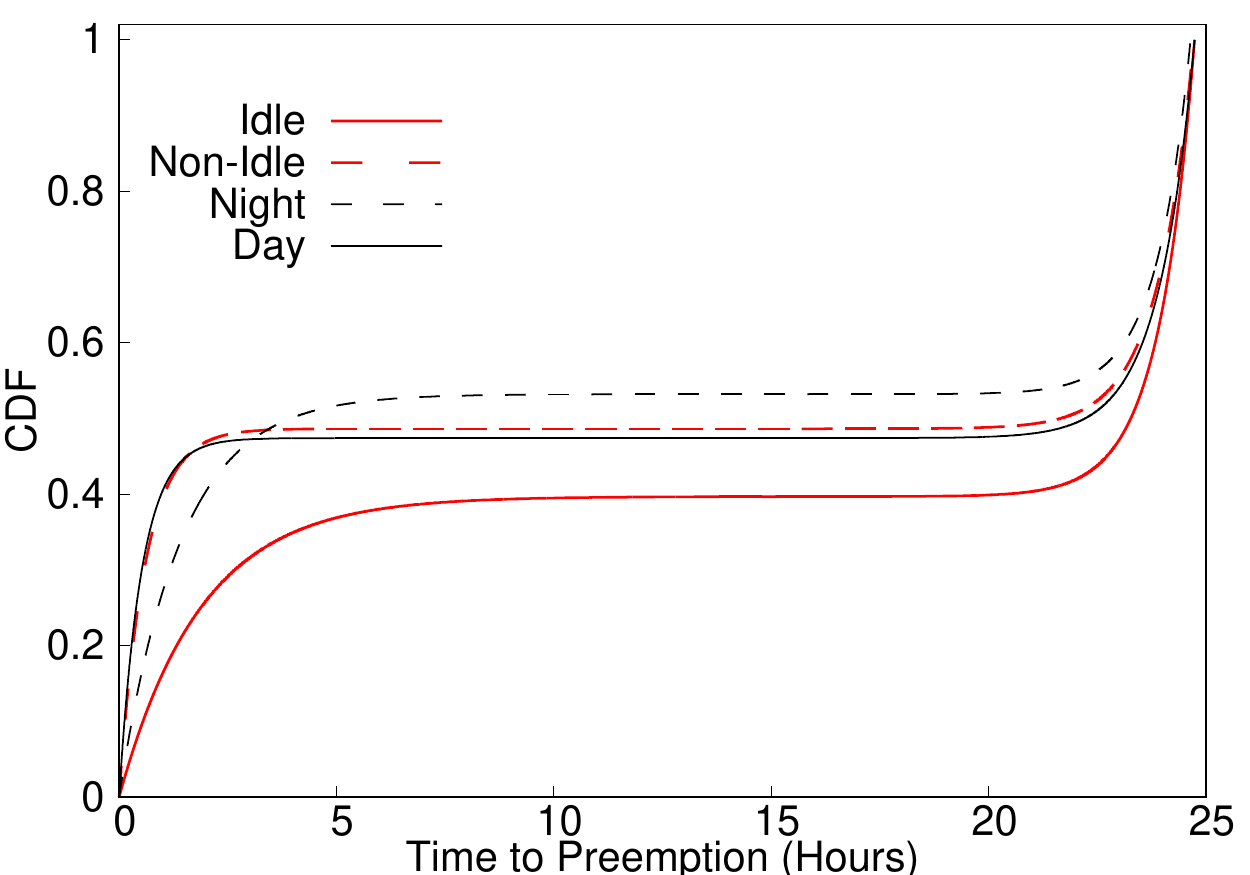} }
\hfill
\subfloat[\textbf{n1-highcpu-16} in different regions. \label{fig:region-breakdown}]
{  \includegraphics[width=0.3\textwidth]{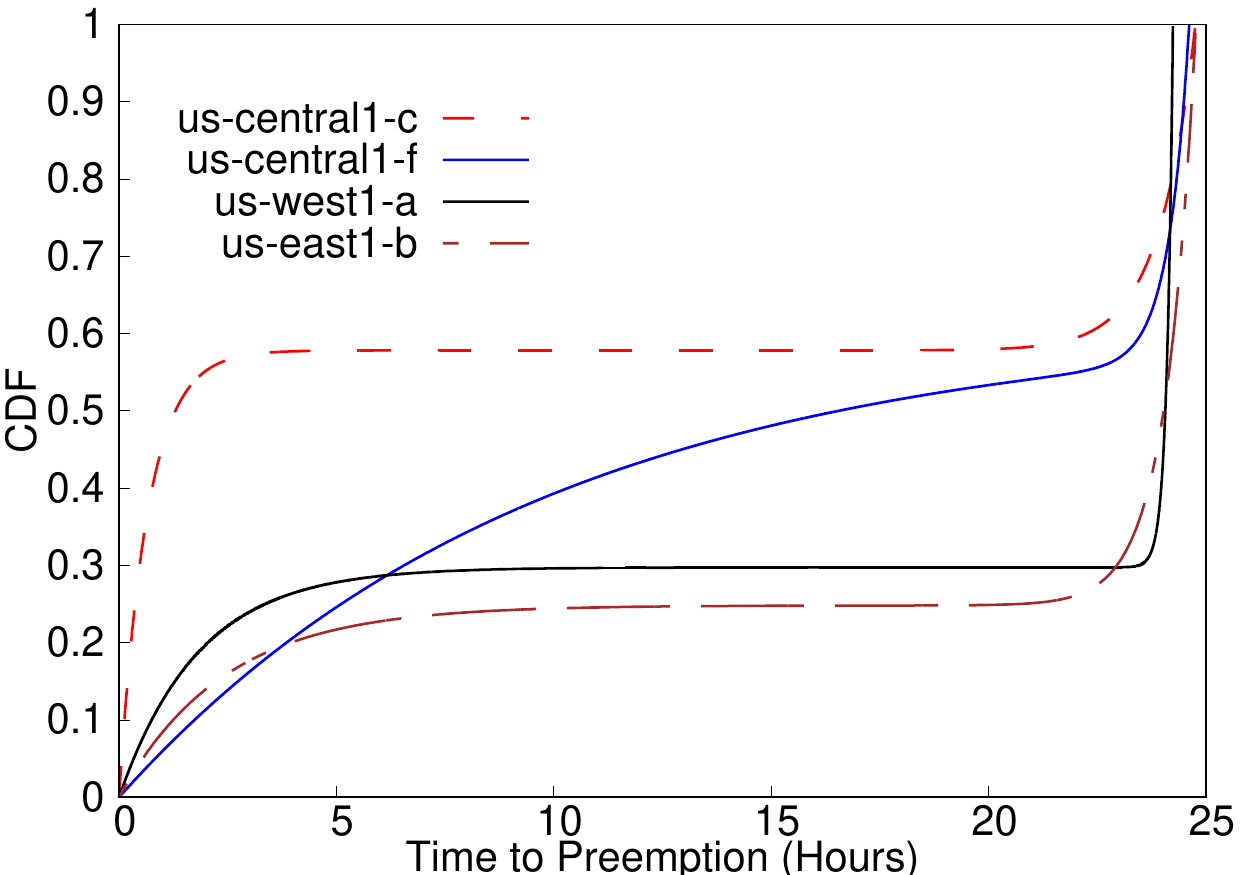} }
\vspace*{-0.6cm}
\caption{Analysis of preemption characteristics by VM-type, region, time-of-day, and workload type.}
\label{fig:breakdown-all}
    \vspace*{\myfigspace}
\end{figure*}

\noindent \textbf{Observation 1:} \emph{The lifetimes of VMs are not uniformly distributed, but have three distinct phases.}

\noindent In the first (initial) phase, characterized by VM lifetime $t\in [0, 3]$ hours, we observe that many VMs are quickly preempted after they are launched, and thus have a steep rate of failure. The rate of failure (preemption rate) is the derivative of the CDF.
The early high rate of failure reflects that the cloud service provider takes into account VM liftetime in prioritizing preempting ``younger'' VMs, in other words, the number of simultaneous VMs launched does have an effect on their failure rate. 
In the second phase, VMs that survive past 3 hours enjoy a relatively low preemption rate over a relatively broad range of lifetime (characterized by the slowly rising CDF in Figure~\ref{fig:gcp1}).
The third and final phase exhibits a steep increase in the number of preemptions as the preemption deadline of 24 hours approaches.
The overall rate of preemptions is ``bathtub'' shaped as shown by the solid black line in the inset of Figure~\ref{fig:gcp1} (discussed in detail below).

\noindent \textbf{Observation 2:} \emph{The preemption behavior, imposed by the constraint of the 24 hour lifetime, is substantially different from conventional failure characteristics of hardware components and EC2 spot instances.}

\noindent In ``classical'' reliability analysis, the rate of failure  usually follows an exponential distribution $f(t) = \lambda e^{-\lambda t}$, where $\lambda=1/\text{MTTF}$.
Figure~\ref{fig:gcp1} shows the CDF ($=1-e^{-\lambda t}$) of the exponential distribution when fitted to the observed preemption data, by finding the distribution parameter $\lambda$ that minimizes the least squares error.
The classic exponential distribution is unable to model the observed preemption behavior because it assumes that the rate of preemptions is independent of the lifetime of the VMs, i.e., the preemptions are \emph{memoryless}.
This assumption breaks down when there is a fixed upper bound on the lifetime. 

\noindent \textbf{Observation 3:} \emph{The three preemption phases and associated bathtub shaped preemption probability are general, universal characteristics of Preemptible VMs.}

Our empirical study looked at preemptions of VMs of different sizes (Figure~\ref{fig:breakdown-all}a), at different times of the day (Figure~\ref{fig:breakdown-all}b), in different geographical zones (Figure~\ref{fig:breakdown-all}c), and running different workloads. 
In all cases, we find that there are three distinct phases associated with the preemption dynamics giving rise to the bathtub shaped preemption probability. 

\noindent \textbf{Observation 4:}\emph{ Larger VMs have a higher rate of preemptions.}

Figure~\ref{fig:cdf-comparison} shows the preemption data from five different types of VMs in the Google Cloud \texttt{n1-highcpu-\{2,4,8,16,32\}}, where the number indicates the number of CPUs.
All VMs are running in the \texttt{us-central1-c} zone. 
We see that the larger VMs (16 and 32 CPUs) have a higher probability of preemptions compared to the smaller VMs.
While this could be simply due to higher demand for larger VMs, it can also be explained from a cluster management perspective. 
Larger VMs require more computational resources (such as CPU and memory), and when the supply of resources is low, the cloud operator can quickly reclaim a large amount of resources by preempting larger VMs.
This observed behavior aligns with the guidelines for using preemptible VMs that suggests the use of smaller VMs when possible~\cite{preemptible-documentation}. 

\noindent \textbf{Observation 5:} \emph{Preemptions exhibit diurnal variations, and are also affected by the workload inside the VM.}

From Figure~\ref{fig:time-breakdown}, we can see that VMs have a slightly longer lifetime during the night (8 PM to 8 AM) than during the day\footnotemark. 
This is expected because fundamentally, the preemption rates are higher during periods of higer demand. 
We also notice that completely idle VMs have longer lifetimes than VMs running some workload.
Presumably, this could be a result of the lower resource utilization of idle VMs being more amenable to resource overcommitment, and result in lower preemptions. 

\footnotetext{Time-zone local to the VM's location.}

\noindent \textbf{Significance of bathtub preemptions.} 
The above empirical observations indicate that temporally constrained preemptions are \emph{not} uniformly distributed. 
The bathtub shaped preemption distribution is not a coincidence.
It is a result of fundamental characteristics of constrained preemptions that benefit applications. 
For applications that do not incorporate explicit fault-tolerance (such as checkpointing), early preemptions result in less wasted work than if the preemptions were uniformly distributed over the 24 hour interval.
Furthermore, the low rate of preemptions in the middle periods allows jobs that are smaller than 24 hours to finish execution with only a low probability of failure, once they survive the initial preemption phase. 
We compare application performance with bathtub preemptions and uniformly distributed preemptions later in Section~\ref{sec:eval}, and find that bathtub preemptions can reduce the performance overheads of preemptions by up $10\times$.
However, effective policies for constrained preemptions  requires a probability model of preemptions, which is challenging due to the temporal constraint and the steep bathtub behavior. 
Existing preemption models are not applicable, and we present our new model next. 

\vspace*{\subsecspace}
\subsection{Failure Probability Model}
\label{subsec:analytical-model}

We now develop an analytical probability model for finding a preemption at time $t$ (preemption dynamics) that is faithful to the empirically observed data and provides a basis for developing running-time and cost-minimizing optimizations. %
Modeling preemptions constrained by a finite deadline raises many challenges for existing preemption models that have been used for other transient servers such as EC2 spot instances.
We first discuss why existing approaches to preemption modeling are not adequate, and then present our closed-form probability model and associated reliability theory connections. 

\vspace*{\subsecspace}
\subsubsection{Inadequacy of existing failure distributions}

Spot instance preemptions have been modeled using exponential distribution~\cite{bid-cloud, hotcloud-not-bid, flint}, which is the default in most reliability theory applications. 
However, the strict 24 hour constraint and the distinct preemption phases are not compatible with the memoryless properties of the exponential distribution. 
To describe failures (preemptions) that are not memoryless (i.e., increasing or decreasing failure rate over time), the classic Weibull distribution with CDF $F(t)=1-e^{-(\lambda t)^k}$ is often employed. However, the Weibull distribution is also unable to fit the empirical data (Figure~\ref{fig:gcp1}) and especially unable to model the sharp increase in preemptions near the 24 hour deadline. 

For constrained preemptions, the increase in failure rate as modeled by the Weibull distribution is not high enough.
Other distributions, such as Gompertz-Makeham, have also been used for modeling bathtub behavior, especially  for actuarial use-cases~\cite{missov2013gompertz}. 
The key idea is to incorporate an exponential aging process, which is used to model human mortality.
The CDF of the Gompertz-Makeham distribution is given by $F(t) = 1 - \exp\left(-\lambda t - \dfrac{\alpha}{\beta}(e^{\beta t} - 1) \right)$
and is fitted to the data in Figure~\ref{fig:gcp1}, and is also unable to provide a good model for the observed preemption data.

The non-trivial bathtub-shaped failure rate (Figure~\ref{fig:gcp1}) requires models that capture the sudden onset of the rise in preemptions near the deadline, which is challenging for the existing failure distributions because of the sharp inflection point. 
From an application and transiency policy perspective, the preemption model must provide insights about the phase transitions, so that the application can adapt to the sharp differences in preemption rates.
For example, the preemption model should be able to warn applications about impending deadline, which existing failure distributions cannot account for. 
Thus, not only is it important to minimize the total distribution fitting error, it is also important to capture the changes in phase.
However,
existing distributions are unable to capture the effects of the deadline and all the phases of the preemptions, and a new modeling approach is needed. 

\vspace*{\subsecspace}
\subsubsection{Our  model}
\label{subsec:preemption-model}

Our failure probability model seeks to address the drawbacks of existing reliability theory models for modeling constrained preemptions. 
The presence of three distinct phases exhibiting non-differentiable transition points (sudden changes in CDF near the deadline, for example) suggests that for accurate results, models that treat the probability as a step function (CDF as a piecewise-continuous function) could be employed.
However, this limits the range of model applicability and general interpretability of the underlying preemption behavior. Our goal is to provide a broadly applicable, continuously differentiable, and informative model built on reasonable assumptions.  

We begin by making a key assumption: the preemption behavior arises from the presence of \emph{two} distinct failure processes.
The first process dominates over the initial temporal phase and yields the classic exponential distribution that captures the high rate of early preemptions.
The second process dominates over the final phase near the 24 hour maximum VM lifetime and is assumed to be characterized by an exponential term that captures the sharp rise in preemptions that results from this constrained lifetime. 

Based on these observations, we propose the following general form for the CDF:

\vspace*{\subsecspace}
\begin{equation}
  \label{eq:blend1}
  \boxed{
  \mathscr{F}\left(t\right) = A\left(1-e^{-\frac{t}{\tau_1}} + e^{\frac{t-b}{\tau_2}}\right)}
  \end{equation}
\noindent where $t$ is the time to preemption, $1/\tau_1$ is the rate of preemptions in the initial phase, $1/\tau_2$ is the rate of preemptions in the final phase, $b$ denotes the time that characterizes ``activation'' of the final phase where preemptions occur at a very high rate, and $A$ is a scaling constant. 
The model is fit to data for $0 < t < L$, where $L \approx 24$ hours represents the temporal interval (deadline).
Combination of the 4 fit parameters ($\tau_1, \tau_2, b$, and $A$) are chosen to ensure that boundary condition $\mathscr{F}(0) \approx 0$ is satisfied.
In practice, typical fit values yield $b \approx 24$ hours, $\tau_1 \in [0.5, 1,5] $, $\tau_2 \approx 0.8$, and $A \in [0.4, 0.5]$.

For most of its life, a VM sees failures according to the classic exponential distribution with failure-rate equal to $1/\tau_1$ -- this behavior is captured by the $1-e^{-t/\tau_1}$ term in Equation~\ref{eq:blend1}. 
As VMs get closer to their maximum lifetime imposed by the cloud operator, they are reclaimed (i.e., preempted) at a high rate $1/\tau_2$, which is captured by the second exponential term, $e^{(t-b)/\tau_2}$ of Equation~\ref{eq:blend1}. 
Shifting the argument ($t$) of this term by $b$ ensures that the exponential reclamation is only applicable near the end of the VM's maximum lifetime and does not dominate over the entire temporal range. 

The analytical model and the associated  distribution function $\mathscr{F}$ introduced above provides a much better fit to the empirical data (Figure~\ref{fig:gcp1}) and captures the different phases of the preemption dynamics through parameters $\tau_1, \tau_2, b$, and $A$. These parameters can be obtained for a given empirical CDF using least squares function fitting methods (we use scipy's \texttt{optimize.curve\_fit} with the dogbox technique~\cite{scipy-fit}). The failure or preemption rate can be derived from this CDF as:
\begin{equation}
  \label{eq:failrate}
    \vspace*{\eqnspace}
f(t) = \dfrac{d \mathscr{F}(t)} {dt} = A \left(\dfrac{1}{\tau_1}e^{-t/\tau_1} + \dfrac{1}{\tau_2}e^{\frac{t-b}{\tau_2}}\right).
\end{equation}
$f(t)$ vs. $t$ yields a bathtub type failure rate function for the associated fit parameters (inset of Figure~\ref{fig:gcp1}).

In the absence of any prior work on constrained preemption dynamics, our aim is to provide an interpretable model with a minimal number of parameters, that provides a sufficiently accurate characterization of observed preemptions data. 
Further generalization of this model to include more failure processes would introduce more parameters and reduce the generalization power. 

\noindent \textbf{Expected Lifetime:} Our analytical model also helps crystallize the differences in VM preemption dynamics, by allowing us to easily calculate their expected lifetime. 
More formally, we define the expected lifetime of a VM ($\mathscr{L}$) as: 
\begin{equation}
  \label{eq:expected-lifetime}
E[\mathscr{L}] =  \int_{0}^{L} t {f}(t)~dt =  -A(t+\tau_1)e^{-t/\tau_1} + A(t-\tau_2) e^{\frac{t-b}{\tau_2}} \biggr\rvert_{0}^{L}
\end{equation}
where $f(t)$ is the rate of preemptions of the VM (Equation~\ref{eq:failrate}).

This expected lifetime can be used in lieu of MTTF, for policies and applications that require a ``coarse-grained'' comparison of the preemption rates of servers of different types, which has been used for cost-minimizing server selection~\cite{flint}. 

\vspace*{\subsecspace}
\section{Application Policies For Constrained Preemptions}
\label{sec:policies}
Having analyzed the statistical behavior of constrained preemptions and presented our probability model, we now examine how the bathtub shape of the failure rate impacts applications. 
Based on insights drawn from our statistical analysis and the model, we develop various policies for ameliorating the effects of preemptions. 
Prior work in transient computing has established the benefits of such policies for a broad range of applications. 
However, the constrained nature of preemptions introduces new challenges that do not arise in other transient computing environments such as Amazon EC2 spot instances, and thus new approaches are required. 
In this section, we first analyze the impact of constrained preemptions on job running time, and then develop new constrained-preemption aware policies for job scheduling and checkpointing. 
We will focus on long-running batch jobs that arise in many applications such as scientific computing. Extensions of our models and policies to distributed applications with different failure semantics is part of our future work. 

\vspace*{\subsecspace}
\subsection{Impact On Running Time}

We now look at how temporally constrained preemptions impact the total expected running time of applications by using our failure probability model. 
When a preemption occurs during the job's execution, it results in wasted work, assuming there is no checkpointing. 
This increases the job's total expected running time, since it must restart after a preemption.
The expected wasted work depends on two factors:
\begin{enumerate} [leftmargin=12pt]
\item The probability of the job being preempted during its execution. 
\item \emph{When} the preemption occurs during the execution. 
\end{enumerate}

We can analyze the wasted work due a preemption using the failure probability model.
We first compute the expected amount of wasted work \emph{assuming} the job faces a single preemption, which we denote by $E[W_1(T)]$, where $T$ is the original job running time (without preemptions).
\begin{equation}
E[W_1(T)] = \int_0^{T} t~P(t | t \leq T)~dt , 
\end{equation}
where $P(t|t\leq T) = P(t) / P(t \leq T)$. Here, $P(t\leq T)$ is the probability that there is a preemption within time $T$ and is given by $P(t \leq T) = F(T)$ where $F(T)$ is the CDF. 
$P(t)$ is the probability of a preemption at time $t$, and is given by $P(t) = f(t)$ , where $f(t)$ is the probability distribution function given by Equation~\ref{eq:failrate}.
We can therefore write the above equation as:
\begin{equation}
  E[W_1(T)] = \int_0^{T} t~P(t | t \leq T)~dt = \frac{1}{F(T)}  \int_0^{T} t~f(t)~dt.
    \label{eq:wasted}
\end{equation}

We note that the integral is the same as the ``expected lifetime'', given by Equation~\ref{eq:expected-lifetime}.
The above expression for the expected waste given a single preemption can be used by users and application frameworks to estimate the increase in running time due to preemptions. 
The total running time (also known as makespan) of a job \emph{with} preemptions is given by:
\begin{equation}
  \label{eq:tot-run-time}
  E[T] = P(\text{no failure})~T + P(\text{1 failure})~\left(T + E[W_1(T)]\right),
\end{equation}
where $P(\text{no failure}) = P(t > T) =  1- F(T)$ and $P(\text{1 failure}) = P(t \leq T) = F(T)$.
Expanding out these terms and using Equation \ref{eq:wasted}, we get
\begin{equation}
  \label{eq:tot-run-time-2}
  E[T] = \left(1-F\left(T\right)\right)T + F(T)\left(T + E[W_1(T)]\right) = T + \int_0^{T} t~f(t)~dt.         
\end{equation}
This expression for the expected running time assumes that the job will be preempted at most once.
An expression which considers the higher order terms and multiple job failures easily follows from the base case, but presents relatively low practical value.

\noindent \textbf{Consequences for applications:}
Based on our analysis, both the increase in wasted time ($E[W_1(T)]/T$) and expected running time $(E[T]/T)$ depend on the length of the job for non-memoryless constrained preemptions. 
For memoryless exponential distributions, the expected waste is simply $T/2$, but this assumption is not valid for constrained preemptions, and thus job lengths must be considered when evaluating the suitability of Preemptible VMs. 

Users and transient computing systems can use the expected running time analysis for scheduling and monitoring purposes.
Since the preemption characteristics are dependent on the type of the VM and temporal effects, this analysis also allows principled \emph{selection} of VM types for jobs of a given length. 
For instance, VMs having a higher initial rate of preemptions are particularly detrimental for short jobs, because the jobs will see high rate of failure and are not long enough to run during the VM's stable period with low  preemption rates. 
We evaluate the expected wasted time and running time for Google Preemptible VMs later in Section~\ref{sec:eval}. 

\vspace*{\subsecspace}
\subsection{Job Scheduling and VM Reuse Policy}

Our bathtub probability model also allows us to develop optimized job-scheduling policies for reducing  job-failures.
Many cloud-based applications and services are \emph{long-running}, and typically run a continuous sequence of tasks and jobs on cloud VMs. 
In the case of deadline-constrained bathtub preemptions, applications face a choice: they can either run a new task on an already running VM, or relinquish the VM and run the task on a \emph{new} VM. 
This choice is important in the case of non-uniform failure rates, since the job's failure probability depends on the ``age'' of the VM. 
Because of the bathtub failure distribution, VMs enjoy a long period of low failure rates during the middle of their total lifespan.
Thus, it is beneficial to \emph{reuse} VMs for multiple jobs, and relinquishing VMs after every job completion may not be an optimal choice. 

However, jobs launched towards the end of VM life face a tradeoff.
While they may start during periods of low failure rate, the 24 hour deadline-imposed sharp increase in preemptions poses a high risk of preemptions, especially for longer jobs.
The alternative is to discard the VM and run the job on a new VM. 
However, since newly launched VMs also have high preemption rates (and thus high job failure probability), the choice of running the job on an existing VM vs. a new VM is not obvious. 

Our job scheduling policy uses the preemption model to determine the preemption probability of jobs of a given length $T$. 
Assume that the running VM's age (time since launch) is $s$.
The intuition is to reuse the VM only if the expected running time is lower, compared to running on a new VM. 
To compute the expected running time of a job of length $T$ starting at vm-age $s$, we  modify our earlier expression for running time (Equation~\ref{eq:tot-run-time-2}) to: 
\begin{equation}
  \label{eq:tot-run-time-s}
    E[T_s]  = T + \int_{s}^{s+T} t~f(t)~dt
  \end{equation}
The alternative is to discard the VM and launch a new VM, in which case the expected running time is $E[T_0]$. Our job-scheduling policy is simple: 
When a job of running time $T$ attempts to start on a VM of age $s$, if $E[T_s] \leq E[T_0]$, then we run the job on the existing VM.
Otherwise, a new VM is launched.
  
This technique can also be used to find the job length $T^*$, when the transition occurs between re-using and launching a new VM.
Thus, accurate job lengths are not necessary, and we only need to know whether $T<T^*$, and only a rough estimate of the job lengths is required. 
We assume that because most scientific computing workloads involve exploration of some parameter space, they often have homogenenous running times. 

\vspace*{\subsecspace}
\subsection{Checkpointing Policy}

A common technique for reducing the total expected running time of jobs on transient servers is to use fault-tolerance techniques such as periodic checkpointing \cite{flint}.
However, the bathtub nature of constrained preemptions requires new checkpointing policies that do not assume memoryless preemptions. 

Checkpointing application state to stable storage (such as network file systems or centralized cloud storage) reduces the amount of \emph{wasted work} due to preemptions.
However, each checkpoint entails capturing, serializing, and writing application state to a disk, and increases the total running time of the application.
Thus, the frequency of checkpointing can have a significant effect on the total expected running time.

Existing checkpointing systems for handling hardware failures in high performance computing, and for cloud transient servers such as EC2 spot instances, incorporate the classic Young-Daly~\cite{dongarra_fault_nodate, daly2006higher, flint, marathe2014exploiting} periodic checkpointing interval that assumes that failures are exponentially distributed and memoryless.  
That is, the application is checkpointed every $\tau = \sqrt{2 \cdot \delta \cdot \text{MTTF}}$ time units, where $\delta$ is the time overhead of writing a single checkpoint to disk.

However, checkpointing with a uniform period is sub-optimal in case of time dependent failure rates, and especially for bathtub failure rates. 
A sub-optimal checkpointing rate can lead to increased recomputation and wasted work, or result in excessive checkpointing overhead. 
Intuitively, the checkpointing rate should depend on the failure rate, and our analytical preemption model can be used for designing an optimized checkpointing schedule.

We now present our checkpointing policy that uses the preemption model and provides non-uniform, failure-rate dependent checkpointing.
In a nutshell, our policy allows us to compute the optimal checkpointing schedule for jobs of different lengths and different starting times, employing a dynamic programming approach that minimizes the total expected makespan. 

\noindent \textbf{Algorithm description:}
Let the uninterrupted running time of the job be $J$.
For ease of exposition, we assume that each job-step takes one unit of time, yielding $J$ job-steps. 
Let the checkpoint cost be $\delta$---i.e, each checkpoint increases the running time by $\delta$. 
We seek to minimize the total expected running time or the \emph{makespan}, which is the sum of $J$, the expected periodic checkpointing cost, and the expected recomputation. 

The makespan $M$ can be recursively defined and computed.
Let $M(J, t)$ denote the makespan where $J$ is remaining length of job to be executed, and $t$ is the time elapsed since the  VM's starting time (i.e., the VM's current age). 
We now need to determine when to take the \emph{next} checkpoint, which we take after $i$ steps. Let $E[M^*]$ denote the minimum expected makespan.
\begin{equation}
  \label{eq:m0}
  E[M^*(J, t)] = \min_{0<i\leq J}{E[M(J, t, i)]}.
\end{equation}
The makespan is affected by whether or not there is a preemption \emph{before} we take the checkpoint: 
\begin{equation}
  \label{eq:m1}
E[M(J, t, i)] = P_{\text{succ}}(t, i+\delta) \cdot E[M_{\text{succ}}] + P_{\text{fail}}(t, i+\delta) \cdot E[M_{\text{fail}}].
\end{equation}
Here $P_{\text{succ}}(t, i+\delta)$ denotes the probability of the job successfully executing without failures until the checkpoint is taken, i.e., from $t$ to $t+i+\delta$. $P_{\text{fail}}(t, i+\delta) = F(t+i+\delta)-F(i+\delta)$ is computed using the CDF, 
and $P_{\text{succ}} = 1 - P_{\text{fail}}$ .

$E[M_{\text{succ}}]$ is the expected makespan if there are no job failures when the job is executing from step $t$ to $t+i+\delta$, and is given by a recursive definition:
\begin{equation}
  \label{eq:msuc}
E[M_{\text{succ}}(J, t, i)] = t+i+\delta + E[M^*(J-i, t+i+\delta)].  
\end{equation}
\noindent The makespan includes the amount of work already done ($t+i$), the checkpointing overhead ($\delta$), and the expected minimum makespan of the rest of the job. 
Similarly, when the job fails before step $i$, then that portion is ``lost work'', and can be denoted by $E[L(t, i+\delta)]$ which is the expected lost work when there is a preemption during the time interval $t$ to $t+i+\delta$.
A preemption before the checkpoint results in no progress, and $J$ steps of the job still remain. 
The expected makespan in the failure case is then given by:
\begin{equation}
  \label{eq:mfail}
 E[M_{\text{fail}}(J, t, i)] = E[L(t, i+\delta)] + E[M^*(J, t+i+\delta)].
\end{equation}

In the case of memoryless preemptions, $E[L(t, i+\delta)]$ is approximated as $\frac{i+\delta}{2}$.
For bathtub preemptions, the lost work is the wasted work that we defined earlier in Equation~\ref{eq:wasted}, but we need to consider the different start and end times, and we get:
\begin{equation}
  \label{eq:exploss}
E[L(t, i+\delta)] = \int_{t}^{t+i+\delta}{x~f(x)~dx}   , 
\end{equation}
where $f(x)$ is the probability density function from Equation~\ref{eq:failrate}.

\noindent \textbf{Computing the optimal checkpoint schedule:}
We can find the minimum makespan $E[M^*(J, t)]$ by using Equations~\ref{eq:m0}--\ref{eq:exploss}. 
Given a job of length $J$, minimizing the total expected makespan involves computing $E[M^*(J, s)]$, where $s$ is the current age of the server. 
Since the makespan is recursively defined, we can do this minimization using dynamic programming, and extract the job-steps at which checkpointing results in a minimum expected makespan. 
The job's checkpointing schedule is determined as follows (assume the job starts at $s=0$ for ease of exposition). 
We first locate the checkpointing interval $i_1$ that minimizes $E[M(J,0,i)]$.  
Then, we recursively find the next checkpointing interval $i_2$ by minimizing  $E[M(J-i_1, i_1,i)]$, and so on, until the $J\leq0$. 

If a job encounters a failure, it is resumed from the most recent checkpoint, on a new VM.
After every such resume-event, we compute the optimal checkpointing schedule for $E[M^*(J_{\text{Remaining}}, 0)]$, since the job's failure rate is dependent on the VM age when it starts, and the job may be resumed at a later time or on a VM of a different type, etc.
Our algorithm yields non-uniform intervals proportional to failure rate.
For a 5 hour job launched on a new VM (time=0), the checkpointing intervals are $(15, 28, 38, 59, 128)$ minutes.
More checkpointing analysis is presented in Section~\ref{subsec:eval-ckpt}. 

\section{Implementing a Batch Computing Service For Preemptible VMs}
\label{sec:impl}

We have implemented a prototype batch computing service that implements various policies for constrained preemptions. 
We use this service to examine the effectiveness and practicality of our model and policies in real-world settings. 
Our service is implemented as a light-weight, extensible framework that makes it convenient and cheap to run batch jobs in the cloud. 
We have implemented our prototype in Python in about 2,000 lines of code, and currently support running VMs on the Google Cloud Platform~\cite{gcp}. 

\begin{figure}[t]
  \includegraphics[width=0.3\textwidth]{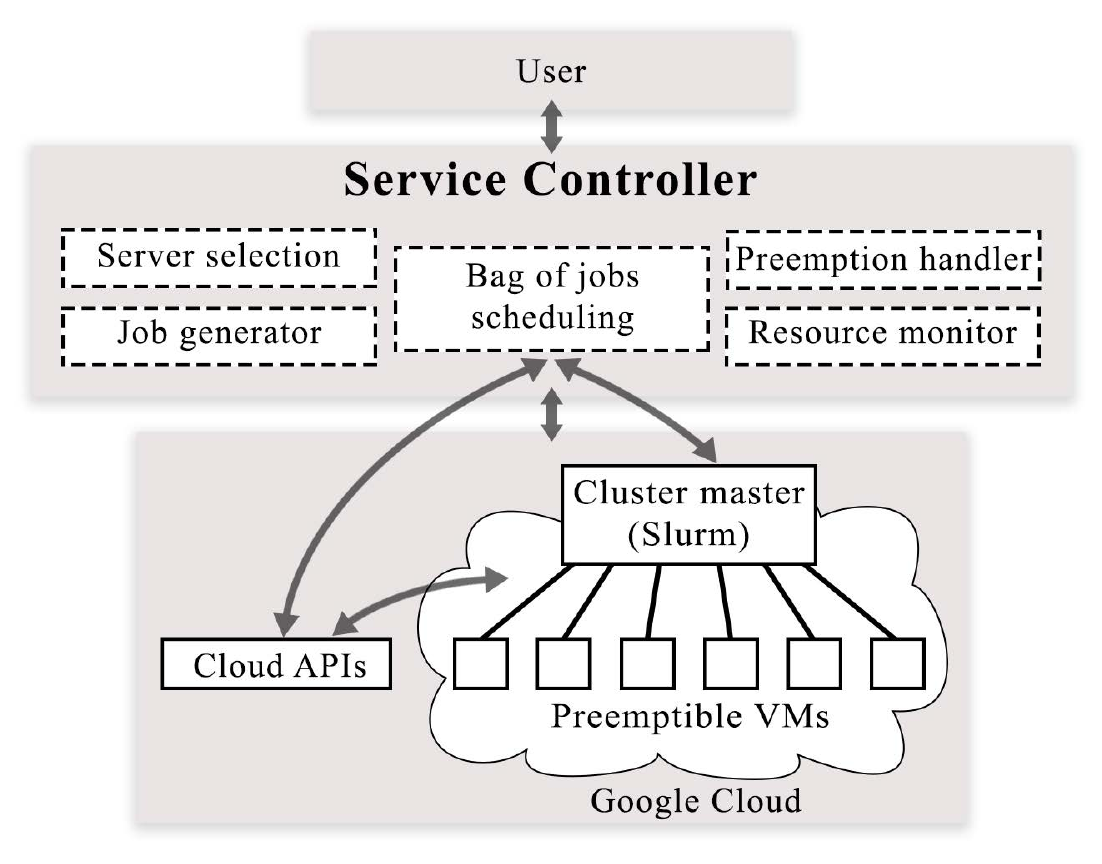}
\vspace*{\myfigspace}
  \caption{Architecture and system components of our batch computing service.}
  \label{fig:arch}
  \vspace*{\myfigspace}
\end{figure}

We use a centralized controller (Figure~\ref{fig:arch}), which implements the VM selection and job scheduling policies described in Section~\ref{sec:policies}. 
The controller can run on any machine (including the user's local machine, or inside a cloud VM), and exposes an HTTP API to end-users. 
Users submit jobs to the controller via the HTTP API, which then launches and maintains a cluster of cloud VMs, and maintains the job queue and metadata in a local database. 

Our service integrates, and interfaces with two primary services.
First, it uses the Google cloud API~\cite{gcloud-api} for launching, terminating, and monitoring VMs.
Once a cluster is launched, it then configures a cluster manager such as Slurm~\cite{slurm} or Torque~\cite{torque}, to which it submits jobs. 
Our service uses the Slurm cluster manager, with each VM acting as a Slurm ``cloud'' node, which allows Slurm to gracefully handle VM preemptions.
The Slurm master node runs on a small, 2 CPU non-preemptible VM, which is shared by all applications and users. 
We monitor job completions and failures (due to VM preemptions) through the use of Slurm call-backs, which issue HTTP requests back to the central service controller. 

\noindent \textbf{Policy Implementation:}
Our service creates and manages clusters of transient cloud servers, manages all aspects of the VM lifecycle and costs, and implements the model-based policies.
It manages a cluster of VMs, and parametrizes the bathtub model based on the VM type, region,  time-of-day, and day-of-week.
When a new batch job is to be launched, we find a ``free'' VM in the cluster that is idle, and uses the job scheduling policy to determine if the VM is suitable or a new VM must be launched. 
Due to the bathtub nature of the failure rate, VMs that have survived the initial failures are ``stable'' and have a very low rate of failure, and thus are ``valuable''.
We keep these stable VMs as ``hot spares'' instead of terminating them, for a period of one hour. 
For the checkpointing policy, our dynamic programming algorithm has a time complexity of $O(T^3)$, for a job of length $T$.
To minimize this overhead, we precompute the checkpointing schedule of jobs of different lengths, and don't need to compute the checkpoint schedule for every new job.

\noindent \textbf{Bag of Jobs Abstraction For Scientific Simulations:}
While our service is intended for general batch jobs, we incorporate a special optimization for scientific simulation workloads that improves the ease-of-use of our service, and also helps in our policy implementation. 
Our insight is that most scientific simulations involve launching a series of jobs that explore a large parameter space that results from different combinations of physical and computational parameters.
These workloads can be abstracted as a ``bag of jobs'', with each job running the same application with  different parameters.
A bag of jobs is characterized by the job and all the different parameters with which it must be executed.
Within a bag, jobs show little variation in their running time and execution characteristics.

We allow users to submit entire bags of jobs, which permits us to determine the running time of jobs based on previous jobs in the bag.
For constrained preemptions, the running time and checkpointing are determined by job lengths, and the job run time estimates are extremely useful. 
Having a large sequence of jobs is also particularly useful with bathtub preemptions, since we can re-use ``stable'' VMs with low preemption probability for running new jobs from a bag.
If jobs were submitted one at a time, a batch computing service may have to terminate the VM after job completion, which would increase the job failure probability resulting from running on new VMs that have a high initial failure rate. 

\vspace*{\subsecspace}
\section{Model and Policy Evaluation}
\label{sec:eval}

In this section, we present analytical and empirical evaluation of constrained preemptions.
We have already presented the statistical analysis of our model in Section~\ref{sec:failmodel}, and we now focus on answering the following questions: 

\begin{enumerate}
\item How do constrained preemptions impact the total running time of applications?

\item  What is the effect of our model-based policies when compared to existing transient computing approaches?

\item What is the cost and performance of our batch computing service for real-world workloads? 
  
\end{enumerate}

\noindent \textbf{Environment and Workloads:}
All our empirical evaluation is conducted on the Google Cloud Platform using our batch computing service described in Section~\ref{sec:impl}. 
All the experiments are conducted in the same time period, and have the same preemption characteristics, as described in our data collection methodology in  Section~\ref{sec:failmodel}. 
We use three scientific computing workloads that are representative of  applications in the broad domains of physics and material sciences:

\noindent \textbf{Nanoconfinement.}
The nanoconfinement application launches molecular dynamics (MD) simulations of ions in nanoscale confinement created by material surfaces \cite{jing2015ionic,kadupitiya2017}. The running time is 14 minutes on a 64 CPU core cluster (4 n1-highcpu-16 VMs). 

\noindent \textbf{Shapes.} The Shapes application runs an MD-based optimization dynamics to predict the optimal shape of deformable, charged nanoparticles \cite{jto1,brunk2019computational}. The running time is 9 minutes on a 64 CPU core cluster (4 n1-highcpu-16 VMs). 

\noindent \textbf{LULESH.} Livermore Unstructured Lagrangian Explicit Shock Hydrodynamics is a popular benchmark 
for hydrodynamics simulations of continuum material models~\cite{IPDPS13:LULESH,LULESH2:changes}. The running time is 12.5 minutes on 8 n1-highcpu-8 VMs.

\vspace*{\subsecspace}
\subsection{Impact of Constrained Preemptions on Job Running Times}

\begin{figure}[t]
  \vspace*{\myfigspace}    
  \subfloat[Computation wasted due to one preemption. \label{fig:vs-uniform}]
  {  \includegraphics[width=0.23\textwidth]{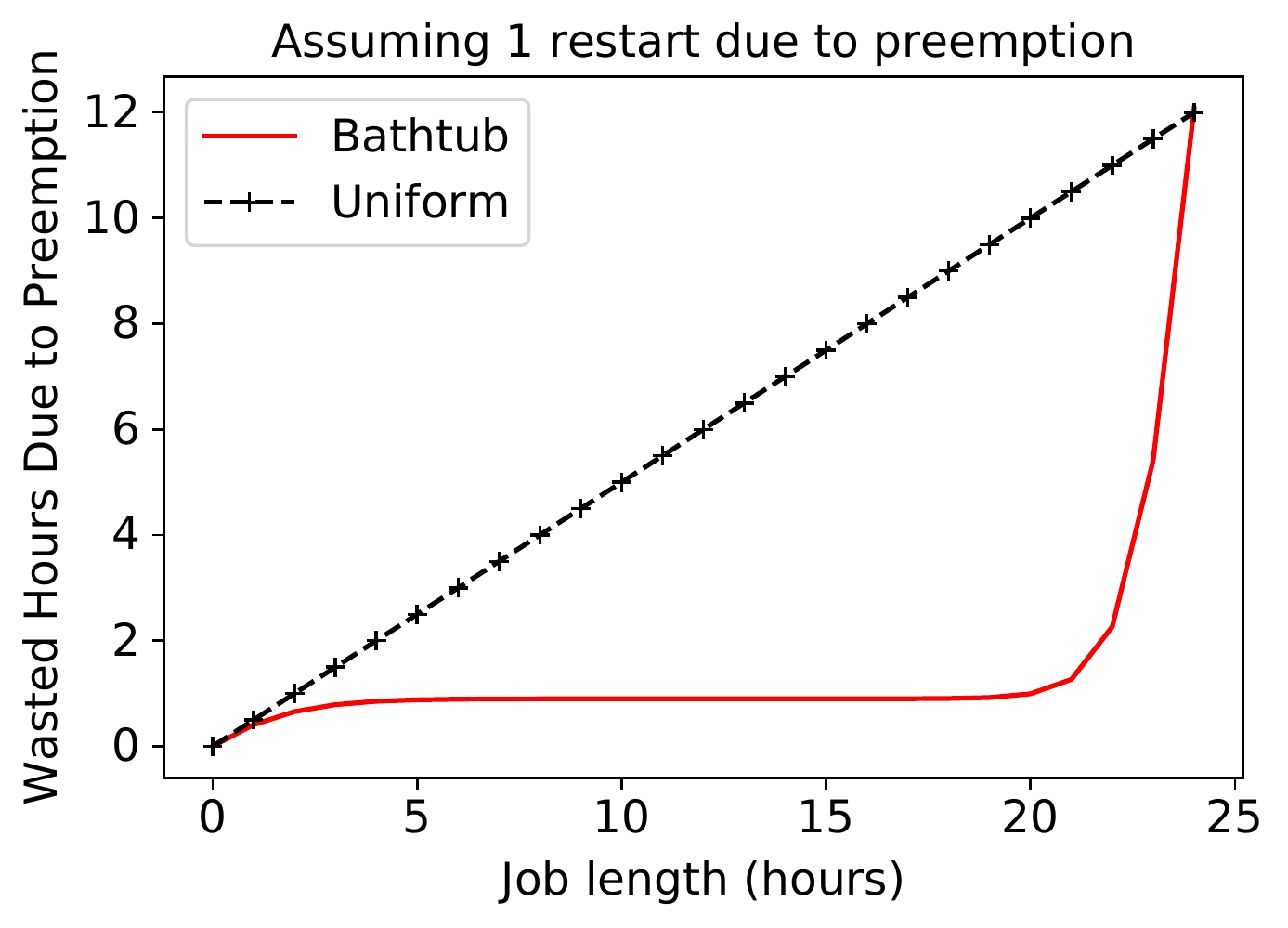} }
    \subfloat[Expected increase in running time. \label{fig:vs-uniform-2}]
    {  \includegraphics[width=0.23\textwidth]{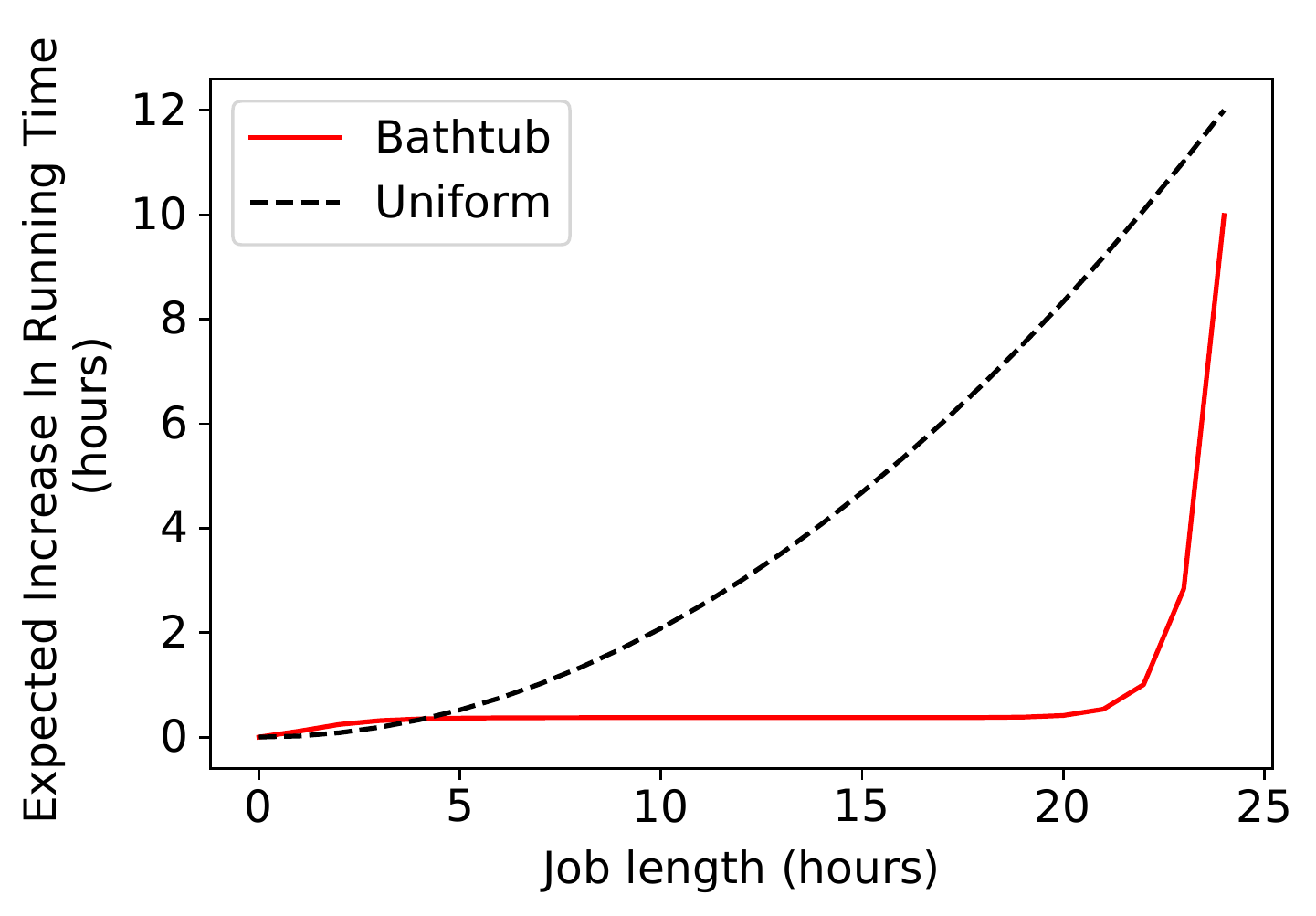} }
    \vspace*{-0.4cm}
    \caption{Wasted computation and expected increase in running time for uniform vs. baththub failures. For jobs $>5$ hours, bathtub distribution results in significantly lower wasted computation.}
\vspace*{\myfigspace}    
  \label{fig:vs-uniform-both}
  
\end{figure}

We begin by examining how constrained preemptions impacts the total job running times. 
When a preemption occurs during the job's execution, it results in wasted work, assuming there is no checkpointing. 
This increases the job's total expected running time, since it must restart after a preemption.
In case of constrained preemptions, the expected waste depends both on the probability of job preemption, as well as \emph{when} the job was preempted. 

For a job of length $J$, the wasted work, assuming that the job faces a \emph{single} preemption, is $E[W_1(J)]$, and is given by Equation~\ref{eq:wasted}.
We first analyze this wasted work for jobs of different lengths in Figure~\ref{fig:vs-uniform}. 
We analyze two failure probability distributions for constrained preemptions: a uniform distribution such that $F(t) = 24-t$, and the bathtub shaped distribution with parameters corresponding to the \texttt{n1-highcpu-16} VM type shown in Figure~\ref{fig:gcp1}. 

For the uniform distribution, the wasted work is linear in the job length, and is given by $J/2$.
For the bathtub distribution, the wasted work is given by Equation~\ref{eq:wasted}.
We now examine the expected increase in running time, that also accounts for the probability of failure, and is given by $P(\text{failure})*E[W_1]$. 
Figure~\ref{fig:vs-uniform-2} shows this expected increase in running times for jobs of different lengths.
We see that for uniformly distributed preemptions, the increase in running time is quadratic in the job length (and is given by $J^2/48$). 
Interestingly, the high rate of early failures for the bathtub distribution results in a slightly worse (i.e., higher) running time for short jobs, because of the high initial rate of bathtub preemptions. 
However for jobs longer than 5 hours, a cross-over point is reached, and the bathtub distribution provides lower overhead of preemptions.
For instance, for a 10 hour job, the increase in running time is about 30 minutes, or 5\%. 
In comparison, if failures were uniformly distributed, the increase would be 2 hours. 

Thus, the bathtub preemptions are beneficial for applications and users, as the low failure rate during the middle periods results in significantly lower wasted work (between $1\times--40\times$), compared to the uniformly distributed failures.
Since the failure rate distribution is ultimately controlled by the cloud provider, our analysis can be used to determine the appropriate preemption distribution based on the job length distributions.
For instance, if short jobs are very common, then uniformly distributed preemptions are preferable, otherwise, bathtub distributions can offer significant benefits. 

\vspace*{\subsecspace}
\subsection{Model-based Policies}
\label{subsec:eval-policy}

We now evaluate the effectiveness of model-driven policies that we proposed earlier in Section~\ref{sec:policies}.
Wherever applicable, we compare against policies designed for EC2 spot instances~\cite{harlap2018tributary, spoton} that have memoryless preemptions. 
However we also note that certain resource management challenges such as the preemption-rate aware job scheduling are \emph{inherent} to constrained preemptions, and no existing equivalent policies can be found for memoryless techniques. 

\vspace*{\subsecspace}
\subsubsection{Job Scheduling}

Previously, we have quantified the increase in running time due to preemptions, but we had assumed that jobs start on a newly launched server.
In many scenarios however, a server may be used for running a long-running sequence of jobs, such as in a batch-computing service. 
Our job scheduling policy is model-driven and decides whether to request a new VM for a job or run it on an existing VM.
A new VM may be preferable if the job starts running near the VM's 24 hour preemption deadline.

\begin{figure*}
  \centering
  \begin{minipage}[c]{0.3\linewidth}
    \includegraphics[width=\linewidth]{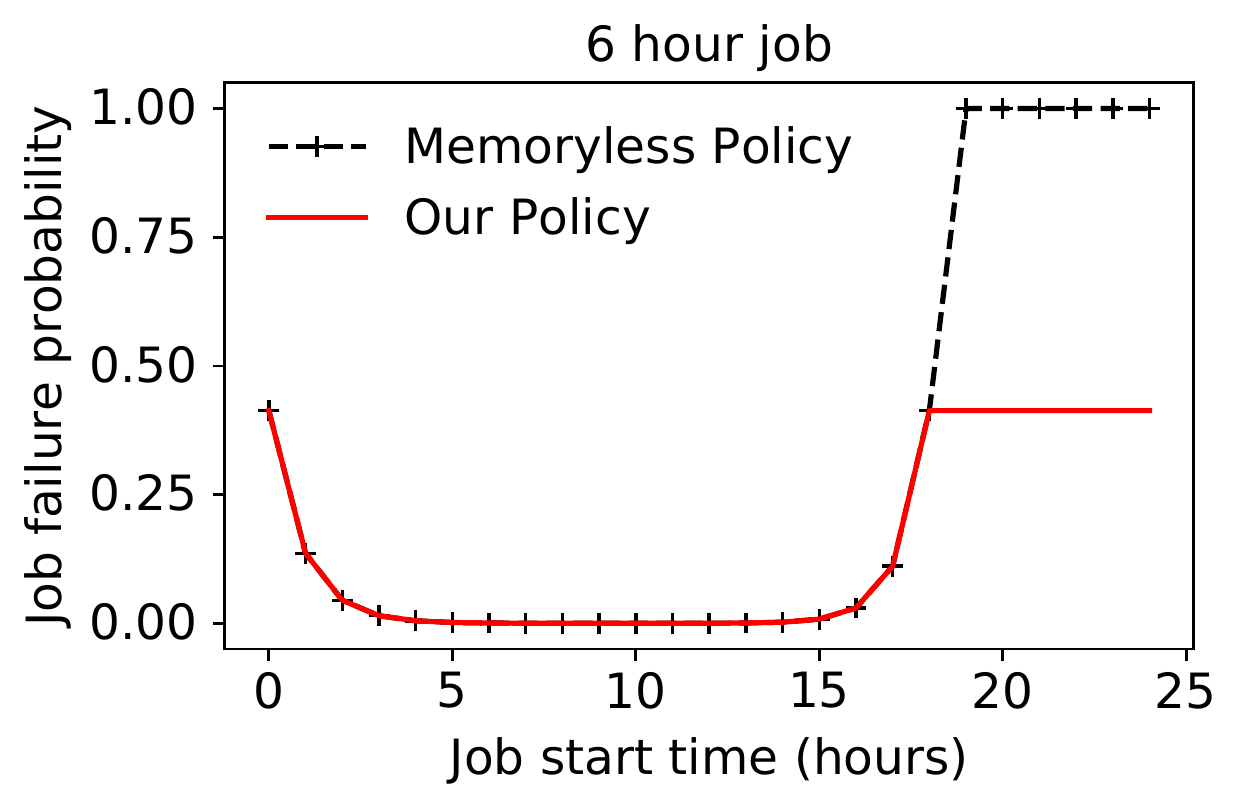}
    \vspace*{-0.6cm}
    \caption{Effect of job start time on the failure probability.} \label{fig:sched-bathtub}
  \end{minipage}
  \hfill 
  \begin{minipage}[c]{0.3\linewidth}
    \includegraphics[width=\textwidth]{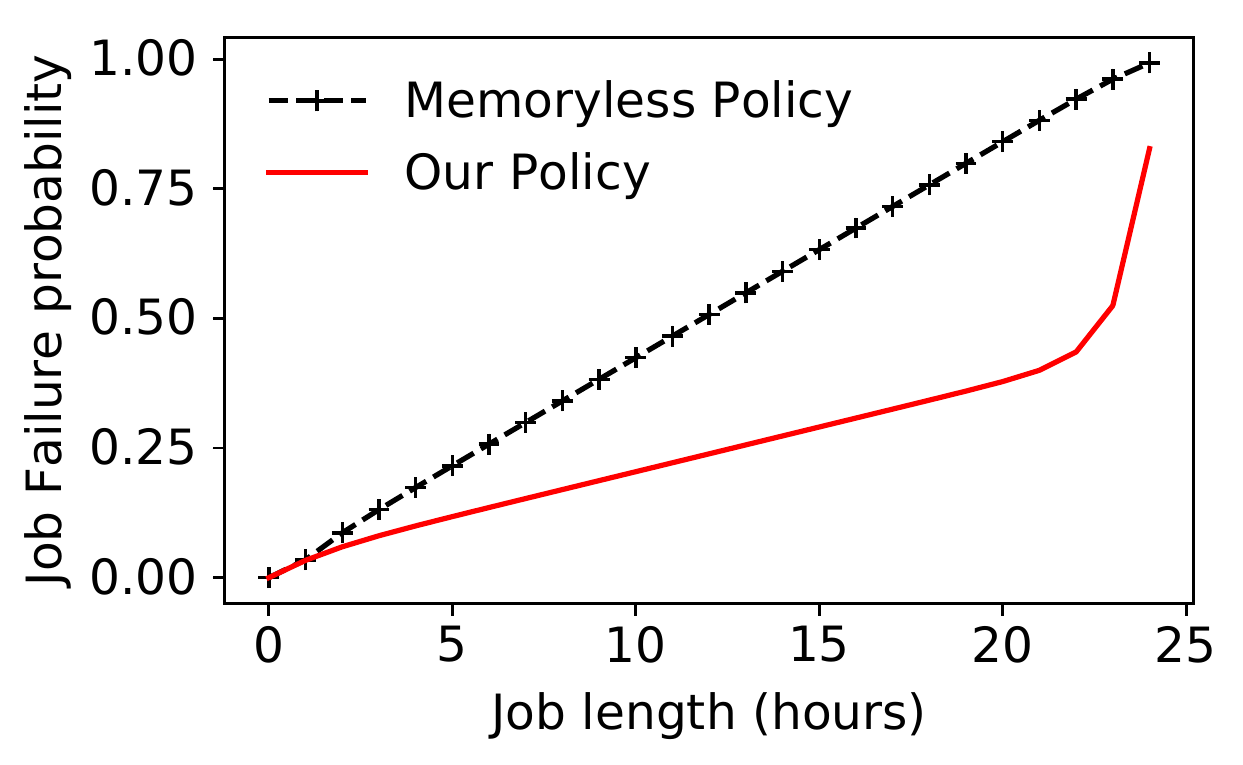}
    \vspace*{-0.6cm}
    \caption{Job failure probability for jobs of different lengths.} \label{fig:sched-all}
  \end{minipage}
  \hfill
  \begin{minipage}[c]{0.3\linewidth}
    \includegraphics[width=\linewidth]{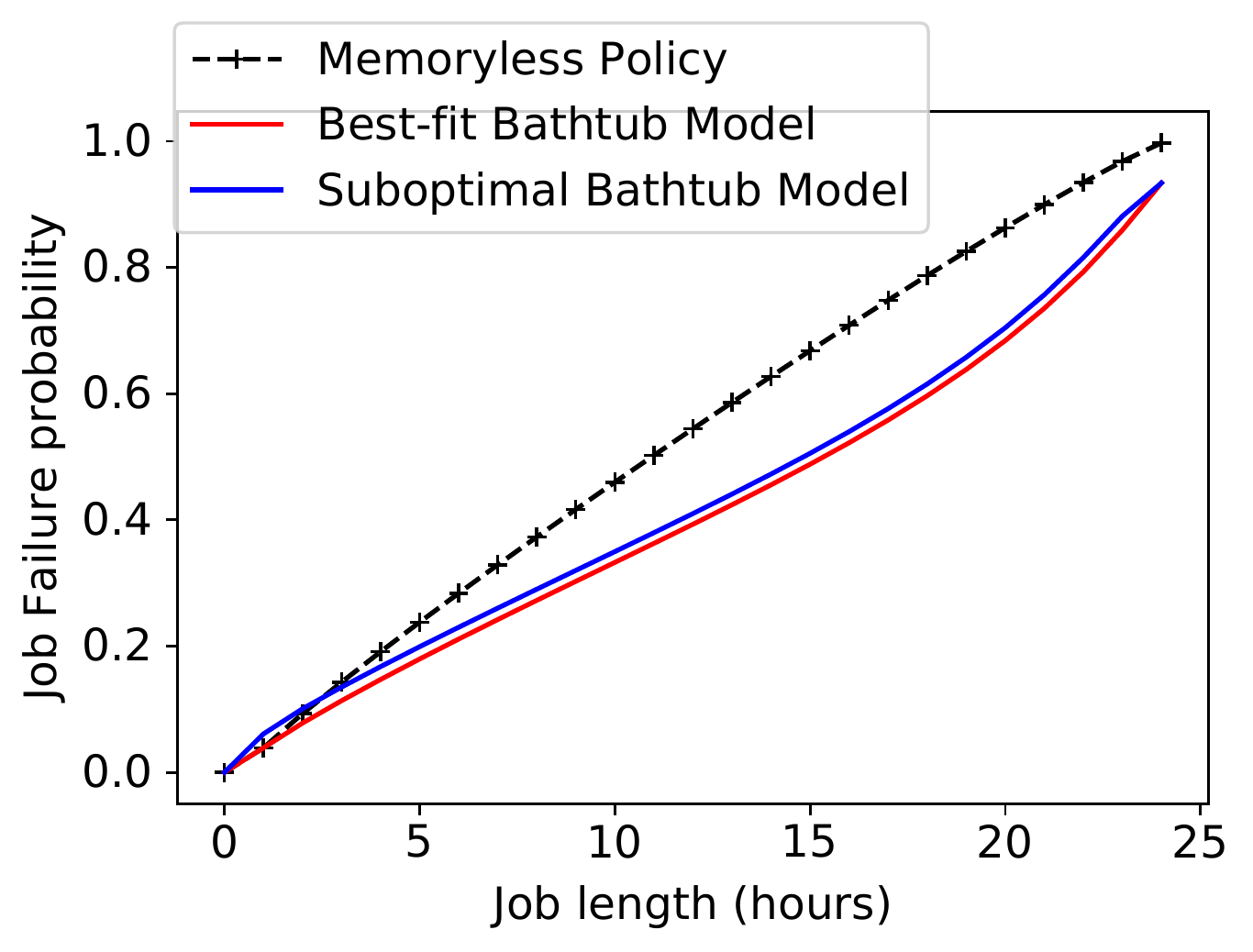}
    \vspace*{-0.8cm}
  \caption{Impact of suboptimal bathtub model parameters on the scheduling policy is negligible.}
  \label{fig:wrong-model}
\end{minipage}
\vspace*{-0.4cm}
\end{figure*}

Figure~\ref{fig:sched-bathtub} shows the effect of our job scheduling policy for a six hour job, for different job starting times (relative to the VM's starting time). 
We compare against a baseline of memoryless job scheduling that is not informed by constrained preemption dynamics.
Such memoryless policies are the default in existing transient computing systems such as SpotOn~\cite{spoton}. 
In the absence of insights about bathtub preemptions, the memoryless policy continues to run jobs on the existing VM. 
As the figure shows, the empirical job failure probability is bathtub shaped. 
However since the job is 6 hours long, with the memoryless policy, it will always fail when launched after $24-6=18$ hours.
In contrast, our model-based policy determines that after 18 hours, we will be better off running the job on a newer VM, and results in a constant lower job failure probability (=0.4). The failure probability is constant because the jobs will always be launched on a new VM after 18 hours, resulting in a failure probability at time=0. 
Thus, our model-based job scheduling policy can reduce job failure probability by taking into account the time-varying failure rates of VMs, which is not considered by existing systems that use memoryless scheduling policies. 

The job failure probability is determined by the job length and the job starting time.
We examine the failure probability for jobs of different lengths (uniformly distributed) in Figure~\ref{fig:sched-all}, in which we average the failure probability across different start times.
We again see that our policy results in significantly lower failure probability compared to memoryless scheduling.
For all but the shortest and longest jobs, the failure probability with our policy is \emph{half} of that of existing memoryless policies. 
This reduction is primarily due to how the two policies perform for jobs launched near the end of the VM preemption deadline, which we examined previously in Figure~\ref{fig:sched-bathtub}. 

\noindent \textbf{Sensitivity to model fitting.}
The effectiveness of any model-based policy depends on the goodness of fit of the preemption model---i.e., how accurately it captures empirical data. 
We now evaluate the impact on our scheduling policy if incorrect/suboptimal model parameters are with high goodness-of-fit ($r^2$) error are used. 
That is, we seek to understand how sensitive our policies are when the underlying preemption behavior does not match the model, which can occur due to changes in supply/demand, minor cloud policy changes, etc. 
Figure~\ref{fig:wrong-model} compares the job failure probability with the optimal bathtub model that best fits the empirical data, and a suboptimal  bathtub model intentionally chosen to have a bad fit. 
Specifically, the suboptimal case models the \texttt{n1-highcpu-16} VMs for \texttt{n1-highcpu-32} VMs, which from Figure~\ref{fig:cdf-comparison} we can see are significantly different.
However even with the suboptimal model, the increase in job failure probability is less than 2\% compared to the best-fit model. 
This negligible difference is due to the fact that even a suboptimal model captures the bathtub shape, and this is enough for the policy to make the ``right'' scheduling decision. 

\noindent \textbf{Result:} \emph{Our policies are not particularly sensitive to the exact model parameters, so long as a bathtub distribution is used. Even a suboptimal bathtub model can reduce failure probability by 15\% compared to the memoryless policy.} 

\vspace*{\subsecspace}
\subsubsection{Checkpointing}
\label{subsec:eval-ckpt}

We now evaluate our model-based checkpointing policy, that uses a dynamic programming approach.
With our policy, the checkpointing rate is determined by the VM's current failure rate.
In contrast, all prior work in transient computing and most prior work in fault-tolerance assumes that failures are exponentially distributed (i.e., memoryless), and use the Young-Daly checkpointing interval.
In the Young-Daly approach, checkpoints are taken after a constant period given by $\tau \propto \sqrt{MTTF}$.
However in the case of constrained preemptions with bathtub distributions, the failure rate is time-dependent and not memoryless.

\begin{figure}[t]
  \vspace*{\myfigspace}
  \subfloat[Checkpointing overhead for different job starting times. \label{fig:ckpt-4}]
{  \includegraphics[width=0.23\textwidth]{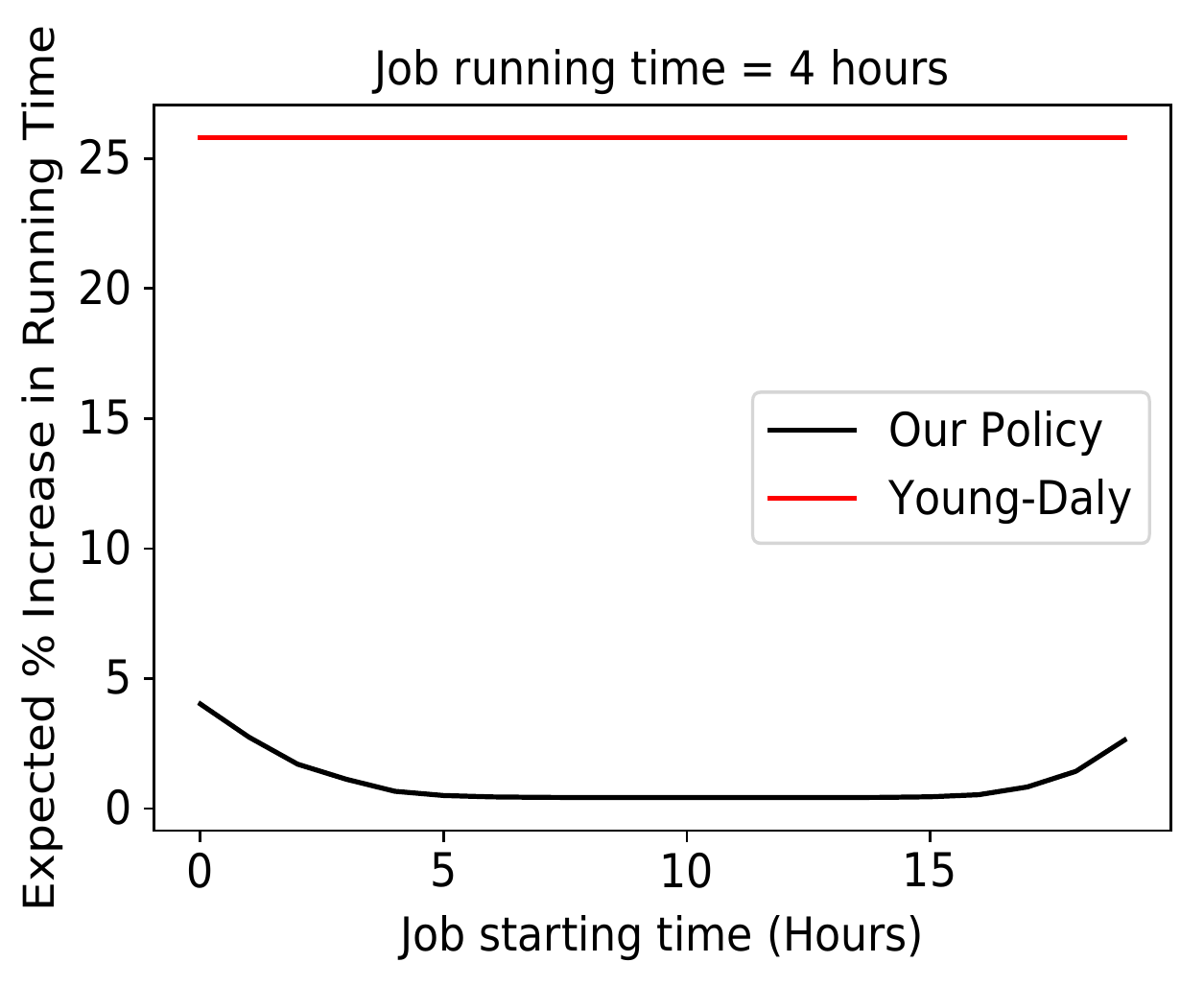} }
\subfloat[Increase in running time with checkpointing when jobs start at time=0. \label{fig:ckpt-start-0-relative}]
{\includegraphics[width=0.23\textwidth]{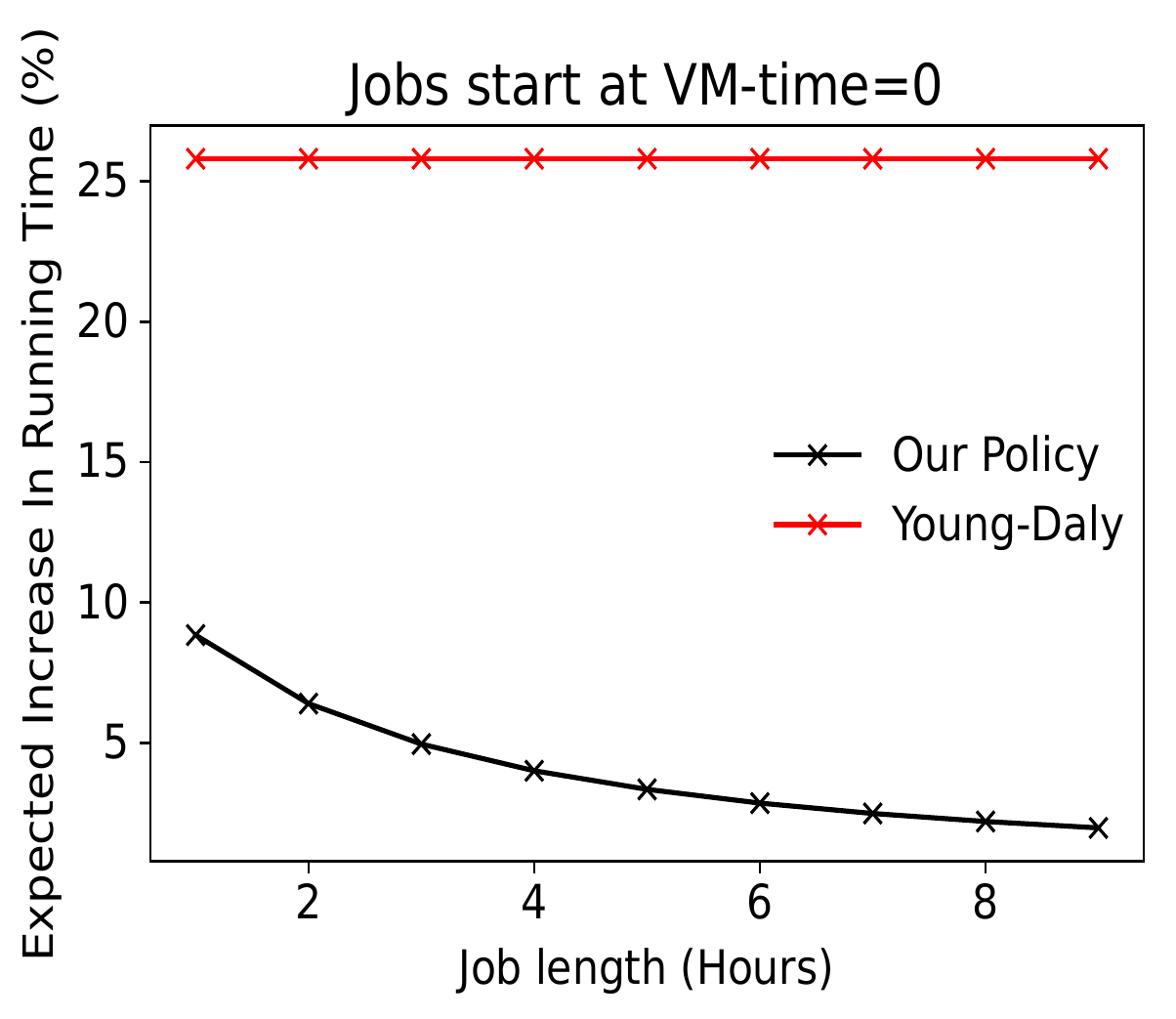}}
\vspace*{\myfigspace}
  \caption{Checkpointing effectiveness.}
  \label{fig:ckpt-all}
  \vspace*{-0.3cm}
\end{figure}

The expected increase in running time for a 4 hour job is shown in Figure~\ref{fig:ckpt-4}, in which we account for both the increase due to the checkpointing overhead, as well as the expected recomputation due to preemptions. 
Throughout, we assume that each checkpoint takes 1 minute. 
The increase in running time depends on the failure rate and thus the job's starting time. 
With our model-based checkpointing policy, the increase in running time is bathtub shaped and is below 5\%, and around 1\% when the job is launched when the VM is between 5 and 15 hours old. 

We also compare with the Young-Daly~\cite{daly2006higher} periodic checkpointing policy, as implemented in~\cite{flint, proteus-eur17, marathe2014exploiting}, and which represents the broad class of fault-tolerance techniques proposed for transient computing. 
For Young-Daly, we use the initial failure rate of the VM to determine the MTTF, which corresponds to an MTTF of 1 hour. 
This results in a high, constant rate of checkpointing, and increases the running time of the job by more than 25\%.
The increase in running time is primarily due to the overhead of checkpointing. 
Note that checkpointing with a lower frequency decreases the checkpointing overhead, but increases the recomputation required.

Next, we examine the expected running time of jobs of different length, when all jobs start at time=0, i.e, are launched on a freshly launched VM.
Figure~\ref{fig:ckpt-start-0-relative} shows the expected increase in the running time of the jobs of length $(0-9] $ hours, with our model-based checkpointing policy and the Young-Daly policy with MTTF=1 hour.
With our policy, the running times increase by 10\% for short jobs ($<2$  hours), and increase by less than 5\% for longer jobs, staying at 3\% on average. 
In contrast, the Young-Daly policy yields a constant increase in running times of 25\%. 
Thus, our model-based policy is able to reduce the checkpointing overhead and thus reduce the performance overhead of running on preemptible VMs to below 5\%. 

\noindent \textbf{Result:} \emph{Our checkpointing policy can keep the performance overhead of preemptions under 5\%, which is $5\times$ better than conventional periodic checkpointing.}

\vspace*{\subsecspace}
\subsection{Effectiveness on Scientific Computing Workloads}

We now show the effectiveness of our batch computing service on Google Preemptible VMs.
We run scientific simulation workloads described earlier in this section, and are interested in understanding the real-world effectiveness of our model-based service.

\begin{figure}[t]
  \vspace*{\myfigspace}
  \centering
  \subfloat[Cost \label{fig:cost-only-bar}]
{  \includegraphics[width=0.25\textwidth]{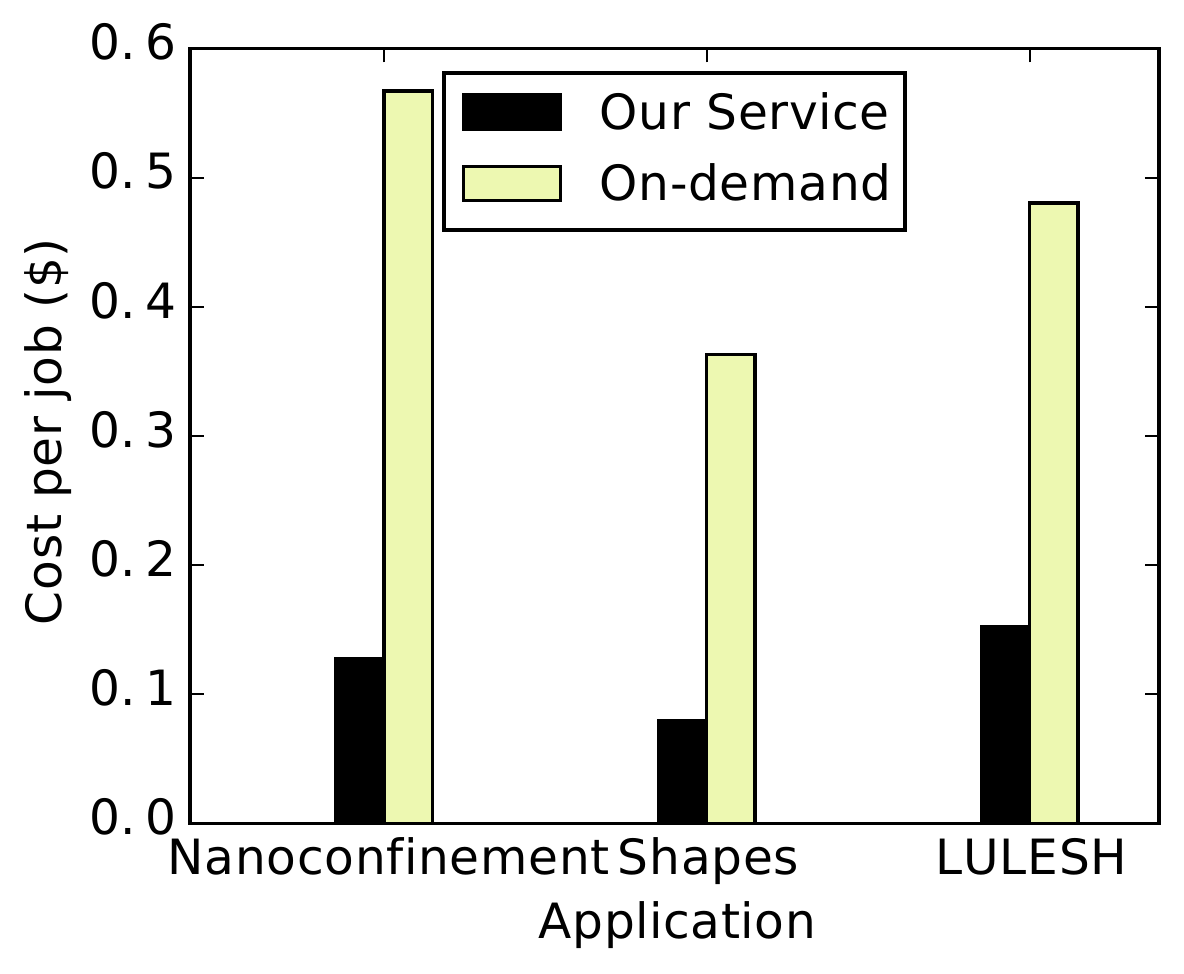} }
\hfill
\subfloat[Preemptions \label{fig:fails-time}]
{ \includegraphics[width=0.2\textwidth]{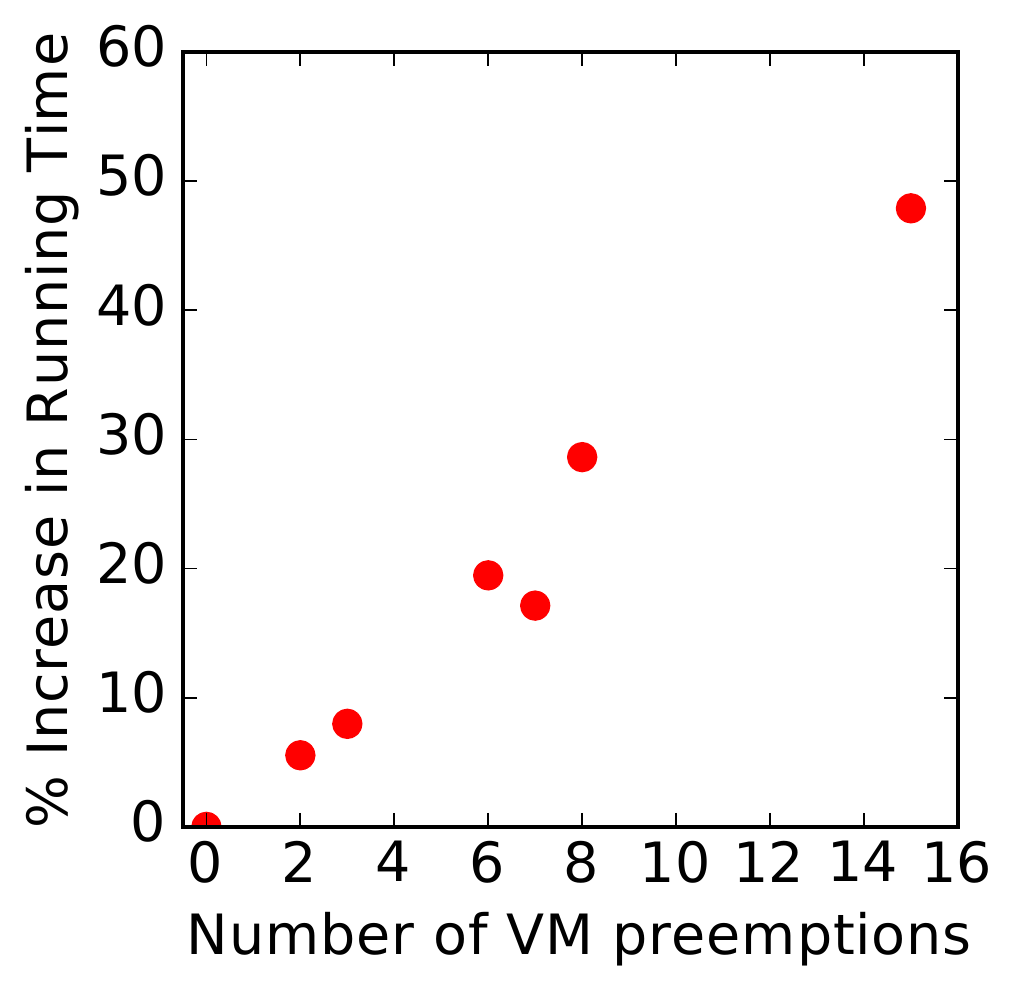} }
\label{fig:service-all}
\vspace*{-0.4cm}
\caption{Cost and preemptions with our service.}
\vspace*{\myfigspace}
\end{figure}  

\noindent \textbf{Cost:}
The primary motivation for using preemptible VMs is their significantly lower cost compared to conventional ``on-demand'' cloud VMs that are non-preemptible.
To evaluate the cost of using our batch computing service, we run a bag of 100 jobs, all running on a cluster of 32 VMs of type \texttt{n1-highcpu-32}. 
Within a bag, different jobs are exploring different physical parameters, and job running times show little variance. 
Figure~\ref{fig:cost-only-bar} shows the cost of using Preemptible VMs compared to conventional on-demand VMs.
We see that our service can reduce costs by $5\times$ for all the applications. 

We note that for this experiment, our  service was using model-driven job scheduling, but was not using checkpointing, since the applications lacked checkpointing mechanisms.
Using checkpointing would reduce the costs even further, since it would reduce the increase in running time (and server costs) due to recomputation.

\noindent \textbf{Preemptions:} 
Finally, we examine the effect of preemptions on the increase in running time under real-world settings.
We ran a cluster of 32 \texttt{n1-highcpu-32} VMs running the Nanoconfinement application, and repeated the experiment multiple times to observe the effect of preemptions.
Figure~\ref{fig:fails-time} shows the increase in running time of the entire bag of jobs, when different number of VM preemptions are observed during the entire course of execution.
We see that the net impact of preemptions results in a roughly linear increase in running time. 
Each preemption results in a roughly 3\% increase in running time, which validates our analytical evaluation shown earlier in Figure~\ref{fig:vs-uniform-2}.
The increase in running time is small because we are computing the \emph{expected} increase in running time, which includes the relatively low probability of preemption. 
The result also highlights the effectiveness of the job scheduling and VM-reuse policy, since most jobs run on the stable VMs, and  those that run on new VMs ``fail fast'' and result in only a small amount of wasted work and increase in running time. 

\noindent \emph{\textbf{Result:} Our batch computing service can reduce costs by up to 5$\times$ compared to conventional on-demand cloud VMs. With the VM-reuse policy, the performance impact of preemptions is as low as 3\%.}

\vspace*{\subsecspace}
\section{Related Work}
\label{sec:related}

\noindent \textbf{Transient Cloud Computing.}
The significantly lower cost of spot instances makes them attractive for running preemption and delay tolerant batch jobs~\cite{spoton, jain14demand, yi2010reducing, conductor, liu-spot, spot-run, dubois2016optispot, varshney_autobot_2019, harlap2018tributary}. 
The challenges posed by Amazon EC2 spot instances, the first transient cloud servers, have received significant attention from both academia  and industry~\cite{spotinst}. 
The distinguishing characteristic of EC2 spot instances is their dynamic auction-based pricing, and choosing the ``right'' bid price to minimize cost and performance degradation is the focus of much of the past work on transient computing~\cite{bidding4,mihailescu2012impact,bidding7,bidding1,bidding8,bidding3,bidding6,bid-cloud,bidding5,wolski_probabilistic_2017, guo_bidding_2015}.
However, it remains to be seen how Amazon's recent change~\cite{bid-change, irwin-icccn19, baughman2019deconstructing, pham2018performance} in the preemption model of spot instances affects prior work. 
Non-price based transient availability models, such as temporally constrained preemptions, have received scant attention due to the difficulty in obtaining empirical preemption data---which we hope our dataset remedies. 
\noindent \textbf{Preemption Mitigation.}
Effective use of transient servers usually entails the use of fault-tolerance techniques such as checkpointing~\cite{flint}, migration~\cite{spotcheck}, and replication~\cite{spoton}. 
In the context of HPC workloads,~\cite{marathe2014exploiting,gong_monetary_2015,xiang_spotmpi:_2011} develop checkpointing and bidding strategies for MPI applications running on EC2 spot instances.
However, periodic checkpointing~\cite{dongarra_fault_nodate, bougeret_checkpointing_2011} is not optimal in our case because preemptions are not memoryless. 

\noindent \textbf{Preemption Modeling.}
Conventionally, exponential distribution have been used to model preemptions, even for EC2 spot instances~\cite{bid-cloud, flint, hotcloud-not-bid}. 
Our preemption model provides a novel characterization of bathtub shaped failure rates not captured even by Weibull distributions, and is distinct from prior efforts~\cite{mudholkar1993exponentiated, crevecoeur1993model}. 
Recent work~\cite{tian_gpu_icdcs} has also found evidence of the bath-tub failure distribution for Google Preemptible GPU VMs, and confirms our observations.
\vspace*{\subsecspace}
\section{Discussion and Future Directions}
\label{sec:discussion}

Constrained preemptions are a relatively unexplored phenomenon and challenging to model.
Our model and the associated data expand transient cloud computing to beyond EC2-spot.
However, many questions and avenues of future investigation remain open:

\noindent \textbf{What if preemption characteristics change?}
Our model allows detecting policy and phase changes by comparing observed data with model-predictions and detect change-points, and 
a long-running cloud service can continuously update the model based on recent preemption behavior. 
However, changes are rare: Google's preemption policy has not changed since its inception in 2015. 
Regardless, VMs with constrained preemptions are an interesting \emph{new} type of transient resource, and our analysis, observations, and policies should continue to be relevant. 
We have also shown that our policies are not particularly sensitive to the model parameters, and even using a ``wrong'' or outdated model can provide significant benefits compared to existing memoryless models. 
Our modeling approach works across a wide range of instance types and is able to model CDFs of instances with both very high and very low failure rates, and thus is general. Moreover, because bathtub preemptions are good for the applications, they will continue to remain a good choice for constrained preemptions making our approach generalizable to other system environments beyond the Google cloud computing systems. Finally, the principle adopted to break down the problem into the superposition of processes characterized by different failure rates can also be considered as a general framework to understand and guide policies for mitigating preemption-induced effects in other cloud environments.

\noindent \textbf{\emph{Phase-wise} model.}
Our statistical analysis indicates that the preemption rates have three distinct phases. 
The analytical model derived in this work is continuously differentiable and allows capturing the three phases reasonably well. 
It may be possible to use a ``phase-wise'' model such as a piece-wise continuously differentiable model, where the three phases are modeled either as segmented linear regions (found using segmented linear regression), or an initial exponential phase and two linear phases. 
Such a piece-wise model could capture the phase transitions with even more accuracy.
Our analytical model informed by empirical data and based on well-defined assumptions can guide the development of such simpler heuristics and modeling approaches, as well as the interpretation of their results.
The analytical model also provides a measure to distill the contributions of effects ignored in its derivation (e.g., only 2 failure rates) in changing the VM expected lifetimes and checkpointing policies. At the same time, it also provides a principled approach to extend the model to more complex cases, e.g., systems characterized by processes with more than 2 failure rates. 

\noindent \textbf{Acknowledgments.} We wish to thank all the anonymous reviewers and our shepherd Ali R. Butt, for  their insightful comments and feedback.
This research was supported by Google cloud credits for research. 
V.J. was partially supported by NSF through Award DMR-1753182.


\end{document}